\documentclass[english,aps,prl, twocolumn]{revtex4-1}
\usepackage[T1]{fontenc}
\usepackage[utf8]{inputenc}
\usepackage{amsmath}
\usepackage{amssymb}
\usepackage{graphicx}
\usepackage{hyperref}
\usepackage{siunitx}
\usepackage{placeins}
\usepackage{braket}
\usepackage{babel}
\usepackage{color}

\newcommand{\Hone}{H^{(1)}}
\newcommand{\Htwo}{H^{(2)}}

\newcommand{\HpSR}[1]{\Hone_{#1}(\kSr R)}
\newcommand{\HpUR}[1]{\Hone_{#1}(\kUr R)}
\newcommand{\HmSR}[1]{\Htwo_{#1}(\kSr R)}
\newcommand{\HmUR}[1]{\Htwo_{#1}(\kUr R)}
\newcommand{\IDR}[1]{I_{#1}(\kaDr R)}

\newcommand{\ceU}{c_{{\rm e} \uparrow}}
\newcommand{\ceD}{c_{{\rm e} \downarrow}}
\newcommand{\chU}{c_{{\rm h} \uparrow}}
\newcommand{\chD}{c_{{\rm h} \downarrow}}
\newcommand{\deU}{d_{{\rm e} \uparrow}}
\newcommand{\deD}{d_{{\rm e} \downarrow}}
\newcommand{\dhU}{d_{{\rm h} \uparrow}}
\newcommand{\dhD}{d_{{\rm h} \downarrow}}

\newcommand{\kFU}{k_{\uparrow}}
\newcommand{\kFS}{k_{\rm S}}
\newcommand{\kSz}{k_{{\rm S}z}}
\newcommand{\kaFD}{\kappa_{\downarrow}}
\newcommand{\kUz}{k_{\uparrow z}}
\newcommand{\kSr}{k_{{\rm S} r}}
\newcommand{\kUr}{k_{\uparrow r}}
\newcommand{\kaDr}{\kappa_{\downarrow r}}

\newcommand{\vUr}{v_{\uparrow r}}

\newcommand{\reeC}{r_{{\rm ee},m}}
\newcommand{\ree}{r_{\rm ee}(k_z,\,\varepsilon)}
\newcommand{\rhoEff}{\rho_{\rm he}}

\newcommand{\me}{m_{\rm e}}
\newcommand{\etaE}{\eta(\varepsilon)}

\renewcommand{\vr}{{\mathbf{r}}}
\newcommand{\OmSxx}{\Omega_{{\rm S}xx}}
\newcommand{\OmSyx}{\Omega_{{\rm S}yx}}
\newcommand{\OmNxx}{\Omega_{{\rm N}xx}}
\newcommand{\OmNyx}{\Omega_{{\rm N}yx}}

\begin{document}

\title{Renormalization effects in spin-polarized metallic wires proximitized by a superconductor: A scattering approach}

\author{Thomas Kiendl}
\email{thomas.kiendl@fu-berlin.de}
\author{Felix von Oppen}
\author{Piet W. Brouwer}
\affiliation{Dahlem Center for Complex Quantum Systems and Fachbereich Physik, Freie Universit\"at Berlin, 14195, Berlin, Germany}

\begin{abstract}
Spin-polarized normal-metal wires coupled to a superconductor can host Majorana states at their ends. These decay into the bulk and are protected by a minigap in the spectrum. Previous studies have found that strong coupling between the superconductor and the normal-metal wire renormalizes the properties of this low-energy phase. Here, we develop a semiclassical scattering approach to explain these renormalization effects. We show that a renormalization of the propagation velocity in the normal wire originates from double Andreev reflection processes at the superconductor interface and that it continues to exist in the absence of a proximity-induced minigap in the normal-metal wire. We also show that the renormalization effects exist for arbitrary transparency of the normal-metal--superconductor interface, provided the superconductor coherence length is sufficiently long in comparison to the thickness of the normal metal.
\end{abstract}

\maketitle

\section{Introduction}

Majorana bound states, zero-energy bound states that are particle-hole symmetric, are predicted to emerge at the ends of one-dimensional topological superconductors. Following theoretical proposals \cite{Lutchyn2010,Oreg2010,chung11,duckheim11, choy11, nadjperge13}, the experimental realization of systems with such Majorana bound states makes use of proximity-induced superconductivity in effectively spin-polarized normal wires, such as a semiconducting wire in a large magnetic field \cite{mourik2012,das2012,churchill13,deng16, albrecht16, chen17, zhang17} or a ferromagnetic wire formed by a chain of magnetic atoms placed on a superconducting substrate \cite{nadjperge14,franke2015,meyer2016,Feldman2017}. In both cases, spin-orbit coupling plays an essential role by allowing the conversion of spin-singlet $s$-wave Cooper pairs in the superconducting substrate into spin-polarized $p$-wave pairs in the proximitized wire.

Not only the zero-energy nature of the Majorana states, but also their localization length can be accessed experimentally. For the atomic-chain platform spatial resolution is a built-in feature of the scanning probe experiment used to detect the Majorana bound state in the first place \cite{nadjperge14,franke2015,meyer2016,Feldman2017}, but spatial information is also available in the semiconductor-wire experiments, by utilizing the hybridization of Majorana states at opposite ends of the wire \cite{albrecht16}. In the atomic-chain experiments, as well as in some of the semiconductor-wire experiments \cite{das2012}, the product of the observed Majorana localization length $l_{\rm maj}$ and the proximity-induced minigap $\varepsilon_{\rm gap}$ was significantly smaller than the expectation $\varepsilon_{\rm gap} l_{\rm maj} \sim \hbar v$ based on models with weak coupling between normal wire and superconductor \cite{dumitrescu15} ($v$ is the Fermi velocity in the normal metal). The anomalously small value of the product $\varepsilon_{\rm gap} l_{\rm maj}$ could be explained by invoking a strong coupling to the superconductor, which substantially renormalizes the properties of the Majorana bound state in atomic chains \cite{Peng_2015, sarma15}, and proximitized semiconductor nanowires \cite{akhmerov17, stanescu17}. The qualitative explanation is that strong coupling to the superconductor places most of the Majorana state's spectral weight in the superconductor, not in the normal metal, which leads to a strong suppression of the propagation velocity along the wire \cite{Peng_2015, sarma15}.

In the present article we consider the velocity renormalization for a spin-polarized wire strongly coupled to a superconductor --- where the spin polarization can be a consequence of the use of half-metallic materials \cite{deGroot1983, Schwarz1986, Park1998, Son2006}, the use of chains of magnetic adatoms \cite{nadjperge14, franke2015, meyer2016, Feldman2017}, or of the application of a magnetic field. The velocity renormalization exists independently of the appearance of a proximity-induced minigap $\varepsilon_{\rm gap}$ in the wire and the possible existence of Majorana bound states.  A strong velocity normalization can exist even if $\varepsilon_{\rm gap}$ is much smaller than the bulk superconducting gap $\Delta$. Such a situation is markedly different from a conventional normal-metal--superconductor junction, where (in the absence of a magnetic field) a large spectral weight inside the superconductor coincides with the short-junction limit for which $\varepsilon_{\rm gap}$ and $\Delta$ are of comparable magnitude.

Our theoretical approach complements Refs.\ \onlinecite{Peng_2015,sarma15}, which used a large tunnel matrix element to model the strong coupling between normal metal and superconductor. Instead, we take a wavefunction approach, and characterize the normal-metal--superconductor interface in terms of its transparency. Then, the strongest coupling naturally appears for an ideal interface with unit transparency. For such an ideal interface, the strong coupling regime appears when $\Delta \ll \hbar v/W$, where $v$ is the Fermi velocity in the absence of coupling to the superconductor and $W$ the transverse dimension of the normal metal. Our method is similar to that of Ref.\ \onlinecite{akhmerov17}, which performs an analysis dedicated to the semiconductor-wire model, and extends previous work on the weak-coupling limit by Duckheim and one of the authors \cite{duckheim11}.

The wavefunction approach allows for an instructive semiclassical picture of the velocity renormalization. In this picture, the renormalization results from a delayed specular reflection of electrons in the normal metal at the superconductor interface, as shown in Fig.\ \ref{fig:setup}. At an ideal normal-metal superconductor interface, this reflection process consists of three stages: (1) An electron incident from the normal metal at angle $\theta$ is transmitted into the superconductor. (2) The transmitted electron is Andreev reflected as a hole. This hole cannot re-enter the spin-polarized normal metal because it has the wrong spin. Instead, it is specularly reflected at the superconductor--normal-metal interface. (3) Finally, the hole is in turn Andreev reflected into an electron, which is subsequently transmitted into the normal metal. Because of the finite penetration length into the superconductor, a delay $\sim 2\hbar/\Delta$ is accumulated in this reflection process. For a normal metal wire of thickness $W$ a distance $2 W \tan \theta$ is traveled between subsequent reflection events within a time $2 W/v \cos \theta$. Thus one obtains the effective velocity
\begin{equation}
  v_x \approx \frac{\Delta}{\hbar} W \tan \theta
  \label{eq:vrenorm}
\end{equation}
in the strong coupling regime $\Delta \ll \hbar v/W$. Note that it is the delay for the {\em normal} reflection that causes the velocity renormalization; the velocity renormalization does not involve processes that lead to Andreev reflection of majority electrons into majority holes or vice versa, which is the cause for the proximity-induced minigap in the normal metal. For a non-ideal interface a second reflection channel, direct specular reflection, is added in parallel to this delayed reflection process. 

Spin-orbit coupling in the normal metal and/or the superconductor enables Andreev reflection of majority electrons into majority holes and a small minigap $\varepsilon_{\rm gap}$ opens up in the spectrum of the normal metal, with Majorana bound states forming at the wire ends. The localization length of the Majorana bound state is $\sim \hbar v_x/\varepsilon_{\rm gap}$, with $v_x$ the renormalized normal-state velocity. The strong renormalization of the velocity $v_x$ in the strong coupling limit leads to a strong renormalization of the product of $\varepsilon_{\rm gap}$ and the Majorana-state localization length. Upon comparing expressions for the weak and strong-coupling limits, we find that it is $\varepsilon_{\rm gap}$ that is renormalized in the strong coupling limit, while the Majorana localization length remains unrenormalized. This is in accordance with the Green function analysis of Refs.\ \onlinecite{Peng_2015, sarma15}. 

The outline of this paper is as follows: In Sec.\ \ref{sec:model} we introduce the model of a spin-polarized metal proximity coupled to a superconductor. In Sec.\ \ref{sec:renormalization} we calculate the dispersion $\varepsilon(k_x)$ for propagating states in the normal wire in the absence of spin-orbit coupling. The renormalized velocity $v_x$ is obtained as $v_x = \hbar^{-1} |d\varepsilon/d k_x|$. Spin-orbit coupling is included in Sec.\ \ref{sec:SOcoupling}, in which we derive the properties of the emerging Majorana bound state for a highly transparent limit and compare the results to the limit of an opaque interface. We conclude in Sec. \ref{sec:conclusion}. To keep the analysis simple, the discussion in the main text is for a two dimensional model. We present results for a three-dimensional setup in the appendix. The results for the two and three-dimensional geometries are qualitatively the same.

\section{Model}\label{sec:model}

\begin{figure}
\includegraphics[width=1\columnwidth]{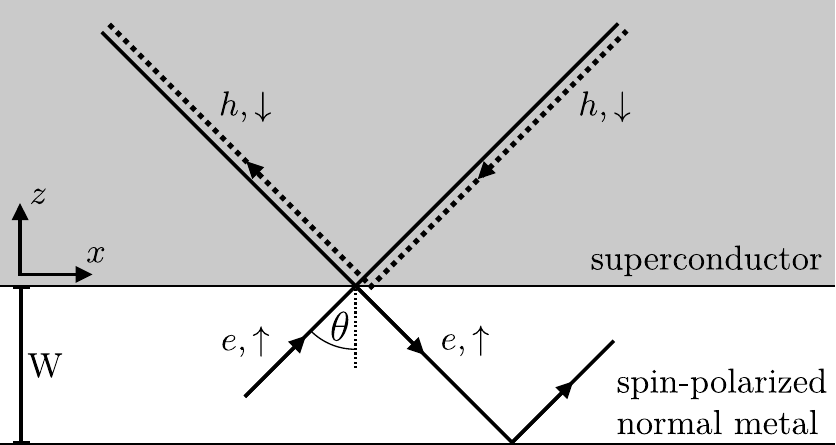}\caption{\label{fig:setup}
Spin-polarized normal-metal wire of width $W$ (white) with one superconducting (grey, top) boundary and one insulating boundary (bottom). In the absence of spin-orbit coupling specular (normal) reflection at the normal-metal--superconductor interface involves a double Andreev reflection process in which an Andreev reflected minority hole is specularly back-reflected into the superconductor. The time delay incurred in this process slows down electrons propagating in the normal metal.}
\end{figure}

We consider a normal-metal (N) strip coupled to a superconductor (S). Coordinate axes are chosen such that the NS interface coincides with the $x$ axis, see Fig.\ \ref{fig:setup}, the superconductor occupies the half space $z > 0$, and the normal metal is in the region $-W < z < 0$. The $4 \times 4$ Bogoliubov-de Gennes (BdG) Hamiltonian reads
\begin{equation}\label{eq:hamiltonian_full}
  \hat{\mathcal{H}} = 
  \begin{pmatrix}
  H_0 & i \sigma_2 \Delta e^{i \phi} \theta(z) \\
  -i \sigma_2 \Delta e^{-i \phi} \theta(z) & - H_0^*
  \end{pmatrix}
\end{equation}
for a BdG spinor $(u_\uparrow,\,u_\downarrow,\,v_\uparrow,\,v_\downarrow)^{\rm T}$ comprising particle and hole wavefunctions. Here $\Delta e^{i \phi}$ is the superconducting order parameter and $\theta(z)$ the Heaviside step function. The $2 \times 2$ normal-state Hamiltonian $H_0$ is
\begin{equation}
  H_0 = \frac{{\bf p}^2}{2m} + V(z) + \frac{\hbar^2 w}{m} \delta(z) + H_{\rm so},
\end{equation}
where $m$ is the electron mass, which we take to be the same in the N and S parts of the system, $V(z)$ is a spin-dependent potential, $(\hbar^2 w/m) \delta(z)$ a potential barrier at the NS interface, and $H_{\rm so}$ the spin-orbit interaction. For the spin-dependent potential $V(z)$, we take different expressions in the normal and superconducting parts of the system,
\begin{equation}
  V(z) = - \frac{\hbar^2 \kFS^2}{2 m}
\end{equation}
when $z > 0$ and 
\begin{equation}
  V(z) = - \frac{\hbar^2}{2 m} \begin{pmatrix}
  k_{\uparrow}^2 & 0 \\
  0 & - \kappa_{\downarrow}^2
  \end{pmatrix} + V_{\rm conf}(z)
\end{equation}
when $z < 0$. Here, $\kFS$ and $\kFU$ are the Fermi wavenumbers of the superconductor and the majority spin band and $V_{\rm conf}(z)$ is a confining potential modeling the sample boundary at $z = -W$, $V_{\rm conf}(z) = 0$ for $z > -W$ and $V_{\rm conf}(z) = \infty$ for $z < -W$. Finally, the spin-orbit coupling is taken to be linear in momentum,
\begin{equation}
  H_{\rm so} = \frac{\hbar}{2} \sum_{j}
  \left[ {\bf p} {\bf \Omega}_j(z) \sigma_j + \sigma_j {\bf \Omega}_j(z) {\bf p}\right],
\end{equation}
where the spin-orbit coupling strength 
\begin{equation}
  {\bf \Omega}_j(z) = {\bf \Omega}_{{\rm S}j} \theta(z) +
  {\bf \Omega}_{{\rm N}j} \theta(-z)
\end{equation}
is piecewise constant in the N and S regions. Spin-orbit coupling is assumed to be weak, so that it can be treated in first-order perturbation theory.

The normal-state majority-carrier transparency of the interface depends on the Fermi velocities $v_{\uparrow} = v = \hbar \kFU/m$ and $v_{\rm S} = \hbar \kFS/m$, the strength $w$ of the surface $\delta$-function potential, and the momentum component $\hbar k_x$ parallel to the interface. In the absence of spin-orbit coupling the corresponding reflection and transmission amplitudes at the Fermi energy $\varepsilon = 0$ are \cite{Kupferschmidt_2011}
\begin{eqnarray} \label{eq:2d_tUp}
  t_{\uparrow}(k_x) &=& \frac{2 \sqrt{k_{\uparrow z} k_{{\rm S}z}}}{2 i w + k_{\uparrow z} + k_{{\rm S}z}},  
  \\ \label{eq:2d_rUp}
  r_{\uparrow}(k_x) &=& -1 + t_{\uparrow}(k_x) \sqrt{k_{\uparrow z}/k_{{\rm S}z}},
  \\ \label{eq:2d_rpUp}
  r'_{\uparrow}(k_x) &=& -1 + t_{\uparrow}(k_x) \sqrt{k_{{\rm S}z}/k_{\uparrow z}},
\end{eqnarray}
where 
\begin{equation}
  k_{\uparrow z} = \sqrt{\kFU^2-k_x^2},\ \
  k_{{\rm S}z} = \sqrt{k_{\rm S}^2 - k_x^2}.
\end{equation}
(The amplitudes $r_{\uparrow}$ and $r'_{\uparrow}$ describe reflection of majority electrons coming from the N and S parts of the system, respectively.) Minority spins coming from $z > 0$ are reflected with reflection amplitude
\begin{eqnarray}\label{eq:2d_rpDown}
  r'_{\downarrow}(k_x) &=& e^{i \varphi_{\downarrow}(k_x)} \nonumber \\
  &=& \frac{k_{{\rm S}z} - i \kappa_{\downarrow z} - 2 i w}{k_{{\rm S}z} + i \kappa_{\downarrow z} + 2 i w},
\end{eqnarray}
where $\kappa_{\downarrow z} = \sqrt{\kappa_{\downarrow}^2 + k_x^2}$ and we neglect terms exponentially suppressed in $\kappa_{\downarrow z} W$.

This model describes semiconductor wires in a large Zeeman field as well as half-metallic (ferromagnetic) wires, both coupled to a superconductor. In the former case spin-orbit coupling is typically assumed to exist inside the semiconductor, but not in the superconductor \cite{Lutchyn2010,Oreg2010}; in the latter case, spin-orbit coupling is usually taken to be in the superconductor, but not in the half-metallic wire \cite{duckheim11,chung11}. 

In the appendix, we consider the corresponding three dimensional model, consisting of a cylindrical spin-polarized normal metal surrounded by a superconductor.

\section{Renormalization of the Fermi velocity} \label{sec:renormalization}

We first consider the system under consideration in the presence of superconductivity, but without spin-orbit coupling. The superconducting gap confines carriers with excitation energy $|\varepsilon| < \Delta$ to the normal region, so that the N region effectively becomes a conducting wire of width $W$. 

Without spin-orbit coupling, reflections at the NS interface are purely normal; Andreev reflections are ruled out because they would require a spin flip process. Nevertheless, the presence of the superconductor can lead to a strong renormalization of the carrier velocity. To see this explicitly, we construct the wavefunction of a majority electron at excitation energy $\varepsilon$ and momentum $\hbar k_x$ parallel to the interface,
\begin{equation} 
  u_{\uparrow}(x,z) \propto e^{i k_x x} 
  \left[ e^{i k_z(k_x,\varepsilon) z} + r_{\rm ee}(k_x,\varepsilon) e^{-i k_z(k_x,\varepsilon) z} \right]. \label{eq:modes_unnormalized_e}
\end{equation}
Here 
\begin{equation}
  k_z(k_x,\varepsilon) = \sqrt{\kFU^2 - k_x^2 + 2 m \varepsilon/\hbar^2}
\end{equation}
and $r_{\rm ee}(k_x,\varepsilon)$ is the reflection amplitude in the presence of the superconductor. In terms of the normal-state reflection and transmission amplitudes of the NS interface the reflection amplitude $r_{\rm ee}(k_x,\varepsilon)$ reads (in the Andreev approximation $\hbar^2 k_z^2/2m \gg \Delta$)
\begin{align}
  r_{\rm ee}(k_x,\varepsilon) &=
  r_{\uparrow}(k_x) + \frac{t_{\uparrow}(k_x)^2 e^{-2 i \eta(\varepsilon) - i \varphi_{\downarrow}(k_x)}}{1 - r'_{\uparrow}(k_x) e^{-2 i \eta(\varepsilon) - i \varphi_{\downarrow}(k_x)}}
  \nonumber \\ &=
  \frac{k_{\uparrow z} - 2 i w - i k_{{\rm S}z} \tan(\eta + \varphi_{\downarrow}/2)}{k_{\uparrow z} + 2 i w + i k_{{\rm S}z} \tan(\eta + \varphi_{\downarrow}/2)}  
, \label{eq:ree_2d}
\end{align}
where 
\begin{equation}
  \eta(\varepsilon) = \arccos(\varepsilon/\Delta).
\end{equation}
This result can be easily understood by considering the different paths a majority electron incident on the NS interface from $z < 0$ can take: Direct normal reflection with amplitude $r_{\uparrow}$ or entering the superconductor with transmission amplitude $t_{\uparrow}$, Andreev reflection into a minority hole, normal backreflection of the hole into S with amplitude $r'^*_{\downarrow}$, finally followed by a second Andreev reflection into a majority electron and transmission into the normal metal. The denominator in Eq.\ (\ref{eq:ree_2d}) describes higher-order processes involving multiple double Andreev reflections. We have assumed $\kappa_{\downarrow} W \gg 1$, so that the minority wavefunction component $u_{\downarrow}$ decays sufficiently fast away from the NS interface and it is sufficient to restrict ourselves to the majority wavefunction component $u_{\uparrow}$.

\begin{figure}
\includegraphics[width=1\columnwidth]{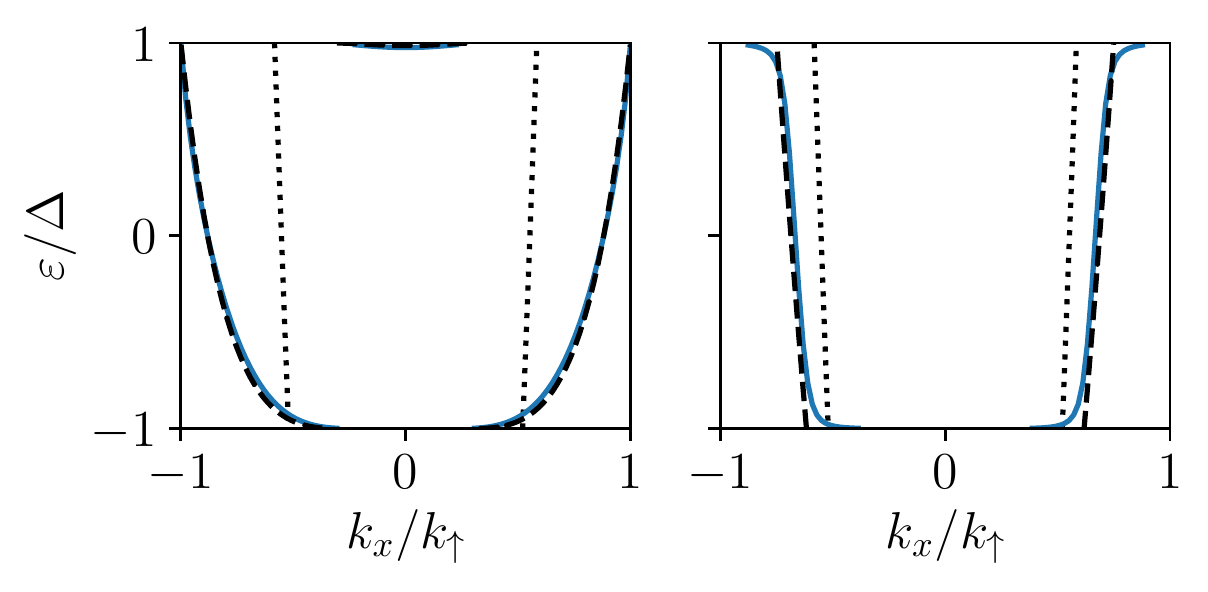}
\caption{\label{fig:dispersion} (Color online.)
Subgap dispersion relation $\varepsilon(k_x)$ for a spin-polarized normal wire attached to a superconductor. Only electron-like solutions are shown, hole-like ones are obtained by mirroring the spectrum vertically such that $\varepsilon\rightarrow -\varepsilon$. The wire width satisfies $\kFU W/\pi = 1.2$, corresponding to one propagating mode at the Fermi level $\varepsilon = 0$ in an isolated wire. The solid lines are obtained by numerically solving Eq. (\ref{eq:2d_non_linear_ev}). The left panel shows the dispersion relation for $\kFU = \kFS$, $w = 0$, corresponding to a fully transparent NS interface; the right panel has $w m/\hbar \kFU = 1$, corresponding to an interface with transmission probability $|t_{\uparrow}|^2 = 1/2$ for perpendicular incidence. The dashed lines show Eqs. (\ref{eq:2d_dispersion_transparent}) (left panel) and (\ref{eq:vxasymp}) (right panel), while the dotted lines show the dispersion for a vanishing interface transparency. The magnitude of the superconducting gap is given by $(\hbar \pi/W)^2/2 m \Delta =10$, well within the validity range of the Andreev approximation. We further set $\kappa_{\rm F \downarrow}/k_{\uparrow} = 2$. }
\end{figure}

The dispersion relation $\varepsilon(k_x)$ follows by imposing that $u_{\uparrow}(x,-W) = 0$, which leads to
\begin{equation}\label{eq:2d_non_linear_ev}	
	1 = -e^{2i k_z W}\ree.
\end{equation}
For a weakly coupled superconductor one has $r_{\uparrow} = r'_{\uparrow} \approx -1$ and $|t_{\uparrow}| \ll 1$, and Eq.\ (\ref{eq:2d_non_linear_ev}) reproduces the standard quantization rule $k_z = n \pi/W$, $n=1,2,\ldots$, and a quadratic dispersion
\begin{equation}
  \varepsilon = \frac{\hbar^2}{2m} \left(
  k_x^2+ \frac{n^2 \pi^2}{W^2} -\kFU^2
  \right).
  \label{eq:free}
\end{equation}
In the opposite limit of an ideal interface with $t_{\uparrow} = 1$ and $r_{\uparrow} = r'_{\uparrow} = 0$, one finds
\begin{equation}
  2 k_z(\varepsilon) W = 2 \eta(\varepsilon) + \varphi_{\downarrow}(k_x) + (2 n+1) \pi.
  \label{eq:kzdisp}
\end{equation}
If we restrict ourselves to the single-mode regime $1 \lesssim \kFU W/\pi \lesssim 2$, the Andreev approximation implies that $(\hbar \pi/W)^2/2m \gg \Delta$, which allows us to neglect the energy dependence on the l.h.s.\ of Eq.\ (\ref{eq:kzdisp}) and obtain the dispersion
\begin{equation}
  \varepsilon = \pm \Delta \sin\left[\frac{\varphi_{\downarrow}(k_x)}{2} - W \sqrt{\kFU^2 - k_x^2}\right]. \label{eq:2d_dispersion_transparent}
\end{equation}
The left panel of Fig.\ \ref{fig:dispersion} shows the dispersion for $\kFU W/\pi = 1.2$ for an ideal interface, together with the approximate result (\ref{eq:2d_dispersion_transparent}) and the dispersion (\ref{eq:free}) of the isolated wire.

Figure\ \ref{fig:dispersion} clearly shows that the coupling to the superconductor leads to significantly flatter $\varepsilon$ vs.\ $k_x$ curves near $\varepsilon = 0$, indicating a strongly renormalized Fermi velocity $v_x = \hbar^{-1}|d\varepsilon/d k_x|$. The strong renormalization of the velocity also follows from the approximate dispersion (\ref{eq:2d_dispersion_transparent}) for an ideal interface,
\begin{eqnarray}\label{eq:2d_renorm_ana}
  v_x &=& \frac{1}{\hbar}\sqrt{\Delta^2 - \varepsilon^2} \frac{k_x W}{\kUz} 
  \left( 1 - \frac{1}{\kappa_{\downarrow z}W} \right). \label{eq:zeroth_order_velocity}
\end{eqnarray}
Although we dropped terms exponentially suppressed in $\kappa_{\downarrow z}W$ in Eq. \ref{eq:2d_rpDown}, we keep the term including $\kappa_{\downarrow z}W$ as it is suppressed by a power law only.
Equation \eqref{eq:2d_renorm_ana} gives an effective velocity $v_x$ that is suppressed by a factor $\Delta/\varepsilon_{\rm kin}$ compared to the velocity $\hbar k_x/m$ of an isolated normal wire. Here, $\varepsilon_{\rm kin} = \hbar^2 \kFU^2/2m$ is the normal-state kinetic energy. This suppression is consistent with the semiclassical estimate (\ref{eq:vrenorm}). 

\begin{figure}
\includegraphics[width=1\columnwidth]{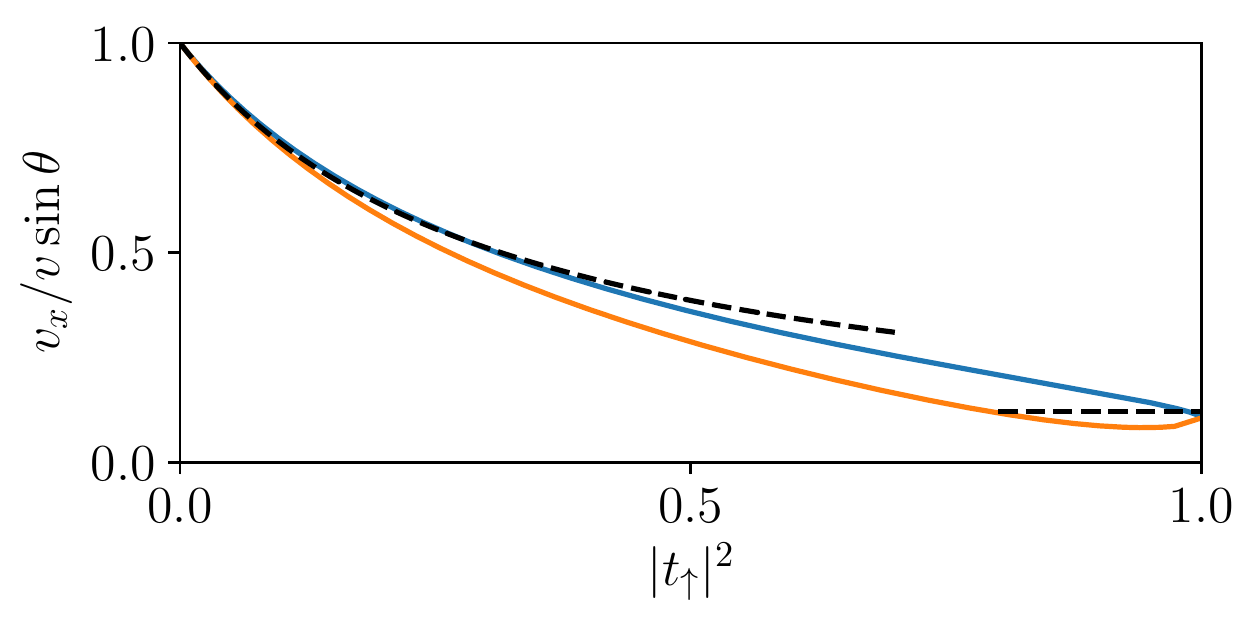}
\caption{\label{fig:renorm_vF_vs_T}Renormalized velocity as a function of interface transparency $|t_{\uparrow}|^2$. The velocity is normalized to $v_x^0 = \hbar k_x/m = v \sin \theta$. The interface barrier is introduced by increasing $w$ while matching $\kFU=\kFS$ (bright, orange line) and by increasing $\kFS$ at fixed $w =0$ (dark, blue line). The solid lines are obtained by numerically solving Eq. (\ref{eq:2d_non_linear_ev}). All other parameters are the same as in Fig. \ref{fig:dispersion}. The dashed lines show the $|t_{\uparrow}|^2=1$ approximation of Eq. (\ref{eq:zeroth_order_velocity}) and the small-transparency approximation of Eq.\ (\ref{eq:vxasymp}).}
\end{figure}

The renormalized velocity is shown in Fig. \ref{fig:renorm_vF_vs_T} as a function of interface transparency for the same parameter choice as in Fig.\ \ref{fig:dispersion}. Starting from the value $v_x = \hbar k_x/m$ of an isolated wire, the velocity decreases monotonically as a function of interface transparency $|t_{\uparrow}|$, reaching the much smaller value given by Eq. (\ref{eq:zeroth_order_velocity}) at $|t_{\uparrow}|^2 = 1$.

Although the velocity renormalization is strongest for a fully transparent interface, we emphasize that the renormalization exists for arbitrary transparency of the interface, provided $\Delta$ is small enough, so that a double Andreev reflection from the superconductor takes a sufficiently long time. In fact, the limit of a weakly transparent interface allows for an explicit solution for $v_x$, as we now show. The limit of a small junction transparency is realized if $k_{{\rm S}z} \gg k_z$ or $|w| \gg k_z$. In this limit one finds
\begin{equation}
  r_{\rm ee} = - \frac{4 w^2 + k_{{\rm S}z}^2 + i k_z (2 w + \varepsilon k_{{\rm S}z}/\Delta)}{4 w^2 + k_{{\rm S}z}^2 - i k_z (2 w + \varepsilon k_{{\rm S}z}/\Delta)},
  \label{eq:ree_limit}
\end{equation}
up to corrections that are small in $|\varepsilon|/\Delta$, in $k_z/|w|$, or in $k_z/k_{{\rm S}z}$. For $|\varepsilon| \ll \Delta$, the solution of Eq.\ (\ref{eq:2d_non_linear_ev}) is 
\begin{equation}\label{eq:2d_kz_lowT_quantization}
  k_z = \frac{\pi}{W} - \frac{\pi (2 w + \varepsilon k_{{\rm S}z}/\Delta)}{W^2(4 w^2 + k_{{\rm S}z}^2)},
\end{equation}
which gives the equation
\begin{equation}\label{eq:2d_eps_low_transparency}
  \varepsilon = \frac{\hbar^2}{2 m}
  \left( k_x^2 + \frac{\pi^2}{W^2} - \frac{2 \pi^2 (2 w + \varepsilon k_{{\rm S}z}/\Delta)}{W^3(4 w^2 + k_{{\rm S}z}^2)} - k_{\uparrow}^2 \right),
\end{equation}
from which the dispersion relation can be obtained. (The $\varepsilon$-dependence of $k_{{\rm S}z}$ can be neglected in the limit of small interface transparency because either $k_{\rm S} \gg k_{\uparrow}$, in which case $k_{{\rm S}z} = k_{\rm S}$ up to small corrections, or $|w| \gg k_{{\rm S}z}$, in which case $k_{{\rm S}z}$ drops out of the equation.) Differentiating with respect to $k_x$ gives the velocity
\begin{equation}
  v_x = \frac{v \sin \theta}{1 + |t_{\uparrow}|^2 \xi_{\rm N}/4 W},
  \label{eq:vxasymp}
\end{equation}
at $\varepsilon = 0$, where $\sin \theta = k_x/k_{\uparrow}$ and $\xi_{\rm N} = \hbar^2 k_z/m \Delta = \hbar^2 \pi/m W \Delta$ is the transverse coherence length in the normal metal. The strong velocity renormalization sets in when $\xi_{\rm N} |t_{\uparrow}|^2 \gg W$. The small-transparency approximation for the dispersion $\varepsilon(k_x)$ and the velocity $v_x$ is illustrated in the right panel of Fig.\ \ref{fig:dispersion} and in Fig.\ \ref{fig:renorm_vF_vs_T}, respectively, showing that the small-transparency approximation remains useful for interface transparencies $|t_{\uparrow}|^2 \lesssim 0.5$.

From a purely classical point of view, the denominator in Eq.\ \eqref{eq:vxasymp} is surprising. To understand this, consider the process shown in Fig.\ \ref{fig:setup} for a low transparency $|t_\uparrow|^2$. From a classical point of view, the electron will spend a time $T_{\rm N} \sim W/v_\uparrow |t_\uparrow|^2$ in the normal metal before being transmitted through the interface and a time $T_{\rm S} \sim  \xi/v_{\rm S} |t_\uparrow|^2$ in the superconducting region. Here, we define the velocities $v_\uparrow = k_\uparrow/m$ and $v_{\rm S} = k_{\rm S} /m$ and neglect the angle $\theta$. In the superconducting region, the distance traveled along $x$ is zero due to the zero-net displacement processes shown in Fig. \ref{fig:setup}, and thus the velocity is expected to be 
\begin{equation}\label{eq:vxasymp_cl}
v_x^{\rm (cl)} \sim \frac{v_\uparrow T_\uparrow}{T_\uparrow + c T_{\rm S}} \sim \frac{v_\uparrow}{1 + c \xi_{\rm N}/W},
\end{equation}
with some constant numerical factor $c$, and the ratio $v_\uparrow/v_{\rm S}$ has been absorbed into $\xi_{\rm N}$. Eq. \eqref{eq:vxasymp_cl} is clearly inconsistent with Eq. \eqref{eq:vxasymp}. The missing factor $|t_\uparrow|^2$ can be traced back to the coherent scattering in the superconductor: During a single cycle of the double Andreev reflection shown in Fig. \ref{fig:cyl_setup}, a phase factor $e^{i \alpha} = e^{-2 i \eta(\varepsilon)} r'_\uparrow (r'_\downarrow)^*$ is picked up. For $|\varepsilon| \ll \Delta$ and a low transparency, this phase factor becomes $e^{i \alpha} = -1 + O(|t|^2)$. Hence multiple double Andreev reflections interfere destructively up to corrections of $O(|t_\uparrow|^2)$ and the time $T_{\rm S}$ is effectively lowered by a factor $|t_\uparrow|^2$, which explains the discrepancy between the classical and semi-classical results in \eqref{eq:vxasymp} and \eqref{eq:vxasymp_cl}.


As shown in the appendix, qualitatively the same results are obtained for a three dimensional setup.

\section{Spin-orbit coupling and Majorana bound states}\label{sec:SOcoupling}

Spin-orbit coupling in the superconductor allows for spin flips and thereby enables Andreev reflections of majority spin electrons into majority spin holes and vice versa. This induces a $p$-wave minigap $\varepsilon_{\rm gap}$ in the excitation spectrum of the normal wire and zero-energy Majorana bound states form at its ends. This section considers both of these effects and relates the localization length $l_{\rm maj}$ of the Majorana bound states and the minigap $\varepsilon_{\rm gap}$ to the renormalization of the Fermi velocity calculated in the previous section. The calculation extends that of Ref.\ \onlinecite{duckheim11}, which considered the same problem in the limit of an opaque NS interface, for which there is no renormalization of the Fermi velocity.

We assume that spin-orbit coupling is sufficiently weak so that it can be treated in first-order perturbation theory. Correspondingly, the probability for Andreev reflection off the normal-metal--superconductor interface is small and the induced minigap $\varepsilon_{\rm gap}$ in the spectrum of the normal wire much smaller than the bulk superconducting gap $\Delta$. For that reason, we neglect corrections to the scattering amplitudes of order $\varepsilon/\Delta$ in the calculations below. 

The starting point of the calculation is an expression for the propagating states in the normal wire in the absence of spin-orbit coupling, normalized to unit flux in the $x$ direction. To keep the notation simple, we restrict to the regime in which there is one propagating mode in the normal-metal wire in the absence of spin-orbit induced Andreev reflection. This mode has transverse wavevector $k_z$, which is determined by the quantization condition (\ref{eq:2d_non_linear_ev}). The electron-like scattering states $|\psi_{{\rm e},\pm}\rangle$ propagating in the positive ($+$) or negative ($-$) $x$ direction have the wavefunction components \cite{Kupferschmidt_2011}
\begin{align}
  u_{\uparrow,\pm}(\vr) =&\, e^{\pm i k_x(\varepsilon) x} \frac{e^{i k_z z} + r_{\rm ee} e^{-i k_z z}}{\sqrt{{\cal N} v_x}} \\
  v_{\downarrow,\pm}(\vr) =&\, -e^{\pm i k_x(\varepsilon) x} \frac{i t_\uparrow \tau_{\downarrow} e^{\kappa_{\downarrow z} z} e^{- i \phi}}{(r'_{\downarrow} + r'_\uparrow)\sqrt{\mathcal{N} v_x}},
\end{align}
in the normal region $-W < z < 0$, where 
\begin{equation}
  k_x(\varepsilon) = \sqrt{k_{\uparrow}^2-k_{\uparrow z}^2} + \frac{\varepsilon}{\hbar v_x},
\end{equation}
with the velocity $v_x$ taken from the calculation of the dispersion in Sec.\ \ref{sec:renormalization}, and
\begin{align}\label{eq:2d_tauDown}
  \tau_{\downarrow} &= \frac{2 \sqrt{k_{{\rm S}z} k_{\uparrow z}}}{ k_{{\rm S}z} + i \kappa_{\downarrow z} + 2 i w}.
\end{align}

Since we are interested in energies $|\varepsilon| \ll \Delta$, we only need to retain the energy dependence in the exponential factors, see the discussion in the previous paragraph. As before, we assume that $\kappa_{\downarrow z} W \gg 1$ so that no hard-wall boundary condition needs to be applied at $z = -W$ for the minority component $v_{\downarrow,\pm}(\vr)$. In the superconducting region, the nonzero wavefunction components are \cite{Kupferschmidt_2011}
\begin{align}
  u_{\uparrow,\pm}(\vr) =& \frac{t_{\uparrow} e^{\pm i k_x(\varepsilon) x - z/\xi}(e^{i \kSz z} - e^{-i \kSz z - i \varphi_{\downarrow}})  }{(1 + r'_{\uparrow} e^{- i \varphi_{\downarrow}}) \sqrt{{\cal N} k_{{\rm S}z} v_x/k_{\uparrow z}}}\nonumber
  , \\
  v_{\downarrow,\pm}(\vr) =& - \frac{i t_{\uparrow} e^{\pm i k_x(\varepsilon) x -z/\xi - i \phi} ( e^{i \kSz z} + e^{-i \kSz z - i \varphi_{\downarrow}} )}{(1 + r'_{\uparrow} e^{- i \varphi_{\downarrow}}) \sqrt{{\cal N} k_{{\rm S}z} v_x/k_{\uparrow z}}}. 
  \label{eq:usuper}
\end{align}
Here 
\begin{eqnarray}
	\kSz &=& \sqrt{\kFS^2 - \kFU^2 + k_z^2},\\
        \xi &=& \frac{\hbar^2 \kSz}{m \Delta}, \\
	\mathcal{N} &=& 2 W + \frac{\mbox{Im}\, r_{\rm ee}}{k_z} +
	\frac{2 \xi_{\rm N} |t_{\uparrow}|^2}{|r'_{\downarrow} + r'_{\uparrow}|^2},
  \label{eq:N}
\end{eqnarray}
where the transverse coherence length in the normal metal $\xi_{\rm N}$ was defined below Eq.\ (\ref{eq:vxasymp}).
The factors $\sqrt{k_{{\rm S}z}/k_{\uparrow z}}$ in the denominators of Eq.\ (\ref{eq:usuper}) are a consequence of current conservation at the normal-metal--superconductor interface.
Similarly, the nonzero wavefunction components of the hole-like scattering states $|\psi_{{\rm h},\pm}\rangle$ are
\begin{align}
  v_{\uparrow,\pm}(\vr) =&\, \frac{e^{\mp i k_x(-\varepsilon) x} (e^{-i k_z z} + r_{\rm ee}^* e^{i k_z z})}{\sqrt{{\cal N} v_x}}, \nonumber \\ 
  u_{\downarrow,\pm}(\vr) =&\, \frac{i t_\uparrow^* \tau_{\downarrow}^* e^{\mp i k_x(-\varepsilon) x} e^{\kappa_{\downarrow z} z}e^{i \phi}}{(r_{\downarrow}'^* + r_\uparrow'^*)\sqrt{\mathcal{N} v_x}}  
\end{align}
in the normal region $-W < z < 0$. Likewise, the corresponding wavefunction components in the superconducting region follow from Eqs.\ (\ref{eq:usuper}) upon exchanging electron and hole components, complex conjugating, and sending $\varepsilon \to -\varepsilon$.

To calculate how spin-orbit coupling modifies these scattering states, we now consider a system for which spin-orbit coupling is non-zero in a segment $0 < x < \delta L$ only. For small enough $\delta L$, spin-orbit coupling induces a backscattering amplitude in the scattering state which is linear in $\delta L$ for small enough $\delta L$. Calculating the linear-in-$\delta L$ scattering amplitudes in perturbation theory in $H_{\rm so}$ as in Ref.\ \onlinecite{duckheim11}, we find for the electron-to-hole amplitude for electrons incident from the left ({\em i.e.}, initially moving in the positive $x$ direction)
\begin{equation}
	\rho_{\rm he} \delta L =  
		- \frac{i}{\hbar} \left\langle \psi_{{\rm h},-} \left|
  \delta \hat{\mathcal{H}}_{\rm so}
		\right| \psi_{{\rm e}, +} \right\rangle,
  \label{eq:matrix_element}
\end{equation}
where 
\begin{equation}
	\delta \hat{\mathcal{H}}_{\rm so} = 
  \frac{1}{2} \left\{ \begin{pmatrix}
		H_{\rm so} & 0 \\
		0 & -H_{\rm so}^*
	\end{pmatrix}, \Theta_{\delta L}(x) \right\},
\end{equation}
with $\{ \cdot, \cdot \}$ the anticommutator and $\Theta_{\delta L}(x) = 1$ for $0 < x < \delta L$ and $\Theta_{\delta L}(x)= 0$ otherwise. This gives 
\begin{align}\label{eq:rho_he}
	\rho_{\rm he} =&\,
        - \frac{i t_{\uparrow}^2 \hbar k_x k_{\uparrow z} (\Omega_{{\rm S}xx} + i \Omega_{{\rm S}yx}) e^{-i \phi} (1 + r_{\downarrow}'^2)}{{\cal N} v_x k_{{\rm S}z}^2 (r'_{\downarrow} + r'_{\uparrow})^2}
        \nonumber 
   		\\ &\, \mbox{} 
   		-
        \frac{
        	2 \hbar k_x (\Omega_{{\rm N}xx} + i \Omega_{{\rm N}yx})
        	t_{\uparrow} \tau_{\downarrow} e^{-i \phi} 
        }{
        	{\cal N} v_{x} (r'_{\uparrow} + r'_{\downarrow})
        	(\kappa_{\downarrow z}^2 + \kUz^2)
        }  \nonumber
		\\
		&\times 
		\left[
			\kappa_{\downarrow z} (1 + r_{\rm ee}) - i \kUz (1 - r_{\rm ee})
		\right]        
        .
\end{align}
The remaining amplitudes are readily obtained by symmetry arguments. The Andreev reflection amplitude $\rho_{\rm he}'$ for incoming electron moving in the negative $x$ direction is obtained from Eq.\ (\ref{eq:rho_he}) by sending $k_x \to -k_x$; The amplitudes for incoming holes are obtained by complex conjugation, $\rho_{\rm eh} = \rho_{\rm he}^*$ and $\rho_{\rm eh}' = \rho_{\rm he}'^*$. Although the wavefunction penetrates a distance $\sim \xi$ into the superconductor, the spatial integrals contributing to the matrix element (\ref{eq:matrix_element}) have support only within a few wavelengths of the interface \cite{duckheim11}. This is the reason why the first term in Eq.\ (\ref{eq:rho_he}) does not involve a factor $\xi$ in the numerator.

The Andreev reflection amplitude $r_{\rm he}(L)$ for a segment of length $L$ can obtained by solving the differential relation \cite{duckheim11}
\begin{equation}\label{eq:rhe_diff_eq}
  \frac{d r_{\rm he}}{d L} = \frac{2 i \varepsilon}{\hbar v_x} + \rho_{\rm he} + \rho_{\rm he}'^* r_{\rm he}^2,
\end{equation}
which is obtained by summing the scattering amplitudes from an infinitesimal slice $0< x<\delta L$ and a subsequent segment $\delta L < x < L$. Integrating Eq. \eqref{eq:rhe_diff_eq} gives the non-perturbative amplitudes
\begin{equation}\label{eq:r_he_eff}
	r_{\rm he}(L) = \frac{\rhoEff \sinh q L}{q \cosh q L - i (\varepsilon/\hbar v_x) \sinh q L}
\end{equation}
and
\begin{equation}\label{eq:r_eh_eff}
	r_{\rm eh}(L) = \frac{\rho_{\rm e h} \sinh q L}{ \cosh q L - i (\varepsilon/\hbar v_x) \sinh q L},
\end{equation}
where 
\begin{equation}
	q = \sqrt{|\rhoEff|^2 - (\varepsilon/\hbar v_x)^2}. \label{eq:q}
\end{equation}
For energies $|\varepsilon| < \varepsilon_{\rm gap}$, with
\begin{equation}
  \varepsilon_{\rm gap} = \hbar v_x |\rho_{\rm he}|
  \label{eq:egap}
\end{equation}
one has $|r_{\rm he}| \to 1$ in the limit $L \to \infty$. This is the hallmark of a Majorana bound state \cite{law09, flensberg10}, with $\varepsilon_{\rm gap}$ being the proximity-induced minigap \cite{duckheim11}. 

With the help of Eq.\ (\ref{eq:q}) one readily identifies $l_{\rm maj} = |\rho_{\rm he}|^{-1}$ as the localization length of the zero-energy Majorana bound state. The strong renormalization of the velocity $v_x$ for a transparent interface enters the denominator of Eq.\ (\ref{eq:rho_he}). However, the fact that in the strong coupling limit $\Delta \ll \hbar v/W$ most of the spectral weight is concentrated in the superconductor also enters into the expression for $\rho_{\rm he}$, through the normalization factor ${\cal N}$. Interestingly, the superconducting gap $\Delta$ drops out from the product ${\cal N} v_x$, causing no additional smallness of the localization length. Nevertheless, the velocity renormalization does affect the product of the minigap and the localization length, in agreement with the analysis of Ref.\ \cite{Peng_2015,sarma15}.

To assess the dependence on interface transparency, it is instructive to evaluate the expressions for the induced gap and the localization length of the Majorana state for a weakly transmitting barrier. Taking the imaginary part of $r_{\rm ee}$ from Eq.\ (\ref{eq:ree_limit}), one concludes that the second term in Eq.\ (\ref{eq:N}) does not contribute to the normalization factor in that limit. Since $|r'_{\downarrow} + r'_{\uparrow}| \simeq 2$ for a weakly transmitting barrier, one finds
\begin{equation}\label{eq:2d_normalization_constant}
  {\cal N} = 2 W + \frac{|t_{\uparrow}|^2 \xi_{\rm N}}{2}.
\end{equation}
To further simplify the expressions for $\rho_{\rm he}$, we consider two special cases: (i) Equal Fermi velocities in the normal metal and the superconductor  $k_{\rm S} = k_{\uparrow}$, and $|w| \gg k_{\uparrow}$ to ensure a non-transparent interface. (ii) $k_{\rm S} \gg k_{\uparrow}$ with a barrier-free interface $w=0$. Here, the small transparency is the result of a large Fermi velocity mismatch between the superconductor and the normal metal. 

In both limits one has $1 + r_{\downarrow}'^2 = 2$, although this equality does not hold generally for non-transparent interfaces. Finally, for the factor $1 + r_{\rm ee}$ we find
\begin{equation}
  1 + r_{\rm ee} = t_{\uparrow}
\end{equation}
in the former limit, and
\begin{equation}
  1 + r_{\rm ee} = - \frac{i t_{\uparrow}^2 \kappa_z}{2 k_{{\rm S}z}}
\end{equation}
in the latter limit (where we assumed that $\kappa_{\downarrow} \ll k_{\rm S}$). For the amplitude whose magnitude is equal to the inverse Majorana localization length, we then find
\begin{align}
  \rho_{\rm he} =&\, i e^{-i \phi} m |t_{\uparrow}|^2 \label{eq:rhohelow1}
  \\ & \mbox{} \times
  \left( \frac{\pi (\Omega_{{\rm N}xx} + i \Omega_{{\rm N}yx})}{\pi^2 + \kappa_{\downarrow z}^2 W^2}
   -
  \frac{\Omega_{{\rm S}xx} + i \Omega_{{\rm S}yx}}{4 \pi}  \right)
  \nonumber
\end{align}
for a weakly transmitting interface with $k_{\rm S} = k_{\uparrow}$ and $|w| \gg k_{\uparrow}$, and
\begin{align}
  \rho_{\rm he} =&\, i e^{-i \phi} m |t_{\uparrow}|^2
  \label{eq:rhohelow2}
  \\ & \mbox{} \times
  \left( \frac{\pi (\Omega_{{\rm N}xx} + i \Omega_{{\rm N}yx})}
  {\pi^2 + \kappa_{\downarrow z}^2 W^2}
   -
  \frac{|t_{\uparrow}|^4
  (\Omega_{{\rm S}xx} + i \Omega_{{\rm S}yx}) }
  {64 \pi}\right)
  \nonumber
\end{align}
in limit of a weakly transmitting interface with $w=0$ and $k_{\rm S} \gg k_{\uparrow}$. Expressions for the induced minigap $\varepsilon_{\rm gap} = \hbar v_x |\rho_{\rm he}|$ follow immediately upon multiplication with the renormalized velocity $v_x$ in Eq.\ (\ref{eq:vxasymp}), restricted to the small-transparency limit.

\begin{figure}
	\includegraphics[width=1\columnwidth]{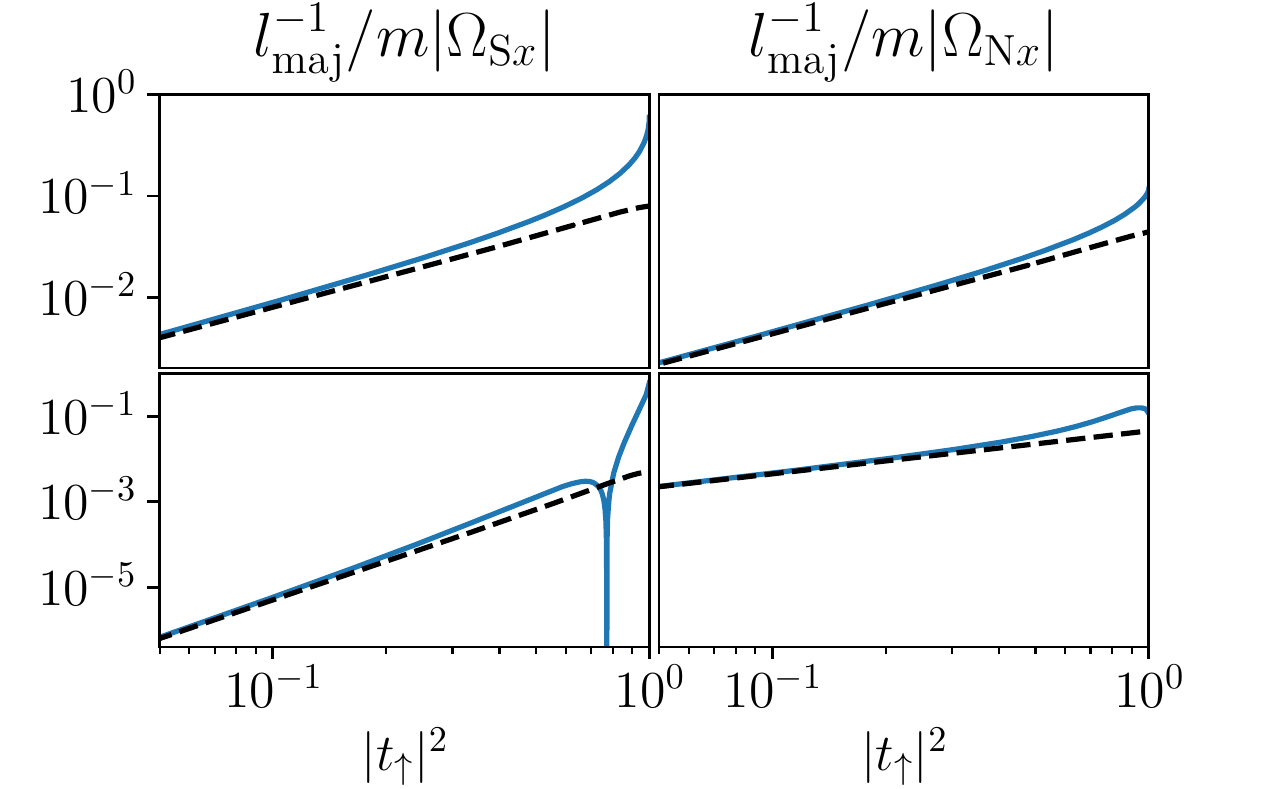}
	\caption{\label{fig:loc-length} Inverse localization length $|\rho_{\rm he}| = 1/l_{\rm maj} $ vs.\ interface transparency $|t_{\uparrow}|^2$ for an interface with matched Fermi velocities $k_{\rm S} = k_{\uparrow}$ (top row) and with zero potential barrier $w=0$ (bottom row), with spin-orbit coupling in the superconductor (left column) and in the normal metal (right column). The dashed curves show the weak-transparency results (\ref{eq:rhohelow1}) and (\ref{eq:rhohelow2}). The remaining parameters are $k_{\uparrow} W = 1.2 \pi$, $(\hbar \pi/W)^2/2 m \Delta = 20$ and $\kappa_{\downarrow} = 2 k_{\uparrow}$. We defined $\Omega_{{\rm S} x}^2 \equiv  {\Omega_{{\rm S}xx}^2 + \Omega_{{\rm S}yx}^2}$ and $\Omega_{{\rm N} x}^2 \equiv {\Omega_{{\rm N}xx}^2 + \Omega_{{\rm N}yx}^2}$.}
\end{figure}

Figure \ref{fig:loc-length} shows the inverse localization length $|\rho_{\rm he}|$ as a function of barrier transparency for the two limits considered above, as well as the full expression \eqref{eq:rho_he} (solid line). For the latter, the velocity and the wave numbers are obtained by numerically solving Eq. \eqref{eq:2d_non_linear_ev}. The figures confirm that the low-transparency expressions in Eqs. (\ref{eq:rhohelow1}) and (\ref{eq:rhohelow2}) are excellent quantitative approximations for transparencies $|t_{\uparrow}|^2 \lesssim 0.5$. However, for transparencies close to unity, spin-orbit coupling in the superconductor, and $w=0$, we observe a sharp closing of the minigap. This is an interference effect which can be traced back to the factor $1 + r_{\downarrow}'^2 = 2 e^{i \varphi_{\downarrow}} \cos \varphi_{\downarrow}$ in Eq.\ (\ref{eq:rho_he}). For $w=0$ and with $\kappa_{\downarrow} > k_{\uparrow}$ the minority reflection phase $\varphi_{\downarrow}$ passes through $\pi/2$ close to unit transparency, see Eq.\ (\ref{eq:2d_rpDown}). A similar effect appears upon approaching perfect transparency by varying $w$ at $k_{\uparrow} = k_{\rm S}$ for negative $w$ (data not shown). 

\begin{figure}
	\includegraphics[width=1\columnwidth]{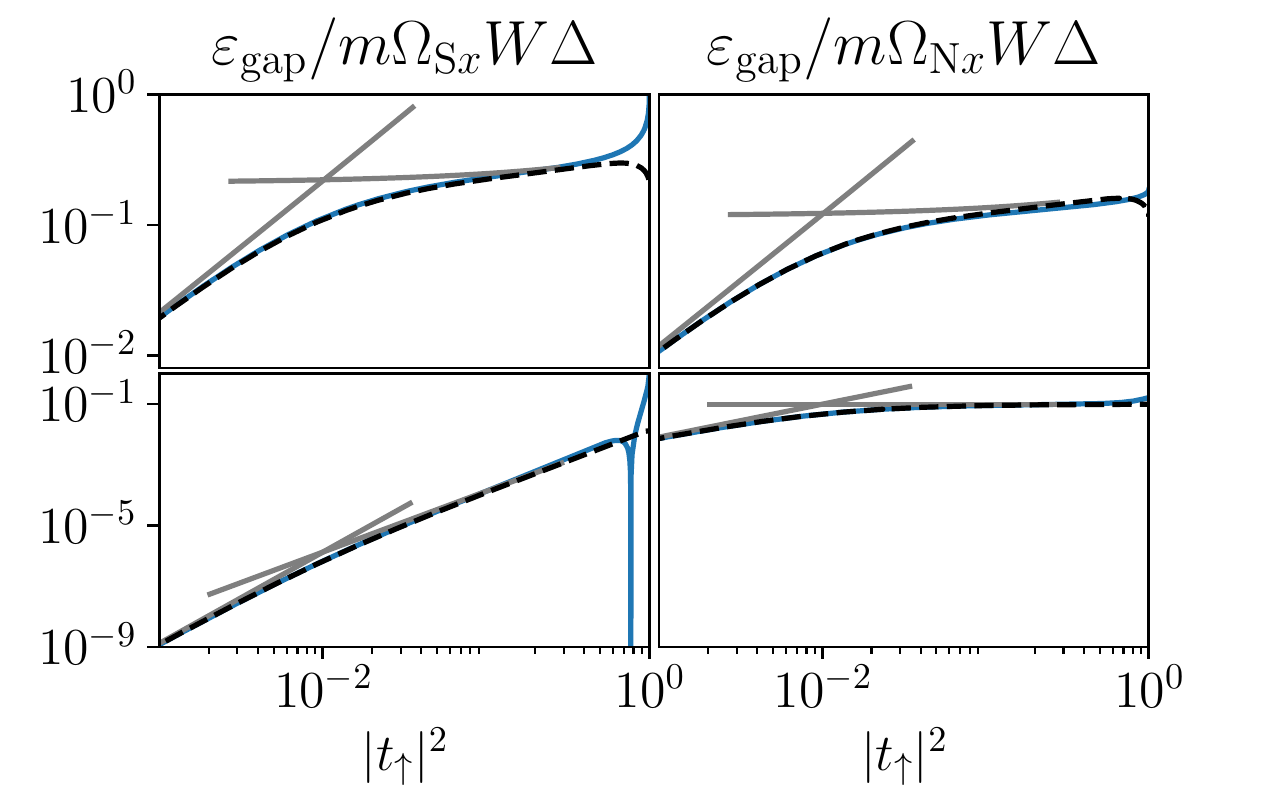}
	\caption{\label{fig:minigap} Minigap versus transparency for the same conditions as in Fig. \ref{fig:loc-length}. The grey curves show the power laws corresponding to the weak-coupling limit $|t_{\uparrow}|^2 \ll W/\xi_{\rm N}$ and the strong-coupling limit (at weak transparency) $W/\xi_{\rm N} \ll |t_{\uparrow}|^2 \ll 1$. The dashed curve is obtained using the weak-transparency results (\ref{eq:rhohelow1}) and (\ref{eq:rhohelow2}) for the inverse localization length $\rho_{\rm he}$. The parameter values are $k_{\uparrow} W = 1.2 \pi$, $(\hbar \pi/W)^2/2 m \Delta = 200 \pi$ and $\kappa_{\downarrow} = 2 k_{\uparrow}$.}
\end{figure}

Figure \ref{fig:minigap} shows the induced minigap $\varepsilon_{\rm gap}$ as a function of barrier transparency. Here the transition between the strong-coupling and weak-coupling limits at $|t_{\uparrow}|^2 \sim W/\xi_{{\rm N}}$ can be clearly seen. The weak-coupling limit agrees with the theory of Ref.\ \onlinecite{duckheim11}; the velocity renormalization appear in the strong-coupling limit $|t_{\uparrow}|^2 \gtrsim W/\xi_{\rm N}$.

\section{Conclusions}\label{sec:conclusion}

In this work, we employed a semiclassical scattering approach to study a spin-polarized normal-metal quantum wire which is strongly coupled to a spin-orbit-coupled superconductor. This model for a topological superconductor was originally introduced and studied in the limit of an opaque interface between wire and superconductor \cite{duckheim11}. Here, we have shown that the properties of its topological phase are strongly renormalized for a highly transparent interface and provide a semiclassical interpretation. Following previous work on related systems \cite{Peng_2015, sarma15, akhmerov17, stanescu17}, we trace the renormalization to the lowering of the Fermi velocity which we interpret in terms of scattering processes which yield zero net-displacement along the wire as well as a modified spin-flip scattering rate $\rhoEff$. Specifically, a transparent interface greatly increases both the topological minigap and the localization length of the emerging Majorana bound states as compared to an opaque one. Additionally we find that, while the low transparency prediction for the localization length stays accurate even for transparencies $\lesssim 0.5$, the velocity as well as the minigap are strongly renormalized towards small values compared to the low-transparency prediction. 

It is interesting to compare our semiclassical approach to the previously employed Green function approach \cite{Peng_2015}. In this approach, one studies the propagation of subgap excitations in the wire, accounting for the coupling to the superconductor through the corresponding self energy
\begin{equation}\label{eq:self_energy}
   \Sigma({\bf k},\omega) = -\Gamma \frac{\omega + \Delta\tau_x}{\sqrt{\Delta^2-\omega^2}}.
\end{equation} 
Here, $\Gamma$ quantifies the coupling between wire and superconductor (with gap $\Delta$) in terms of the decay rate of subgap excitations of the wire (with energy $\omega$) into the superconductor in the normal state. The self energy is written in Nambu notation with the corresponding Pauli matrices denoted by $\tau_j$ ($j=x,y,z$) and does not yet account for spin-orbit coupling in the superconductor. Thus the pairing terms $\propto \tau_x$ describe conventional s-wave pairing and the induced $p$-wave pairing involves a dimensionless measure of the spin-orbit coupling in addition. 

The expression in Eq.\ \eqref{eq:self_energy} is independent of the wave vector ${\bf k}$, making the self energy local in real space. Within the semiclassical picture of the present paper, this surprising locality has a natural interpretation in terms of the locality of the scattering processes by the superconductor. Moreover, the semiclassical approach requires a purely spectral description of the renormalizations. The expression in Eq.\ \eqref{eq:self_energy} implies that we can expect such a spectral interpretation in the limit in which $\omega\ll\Delta$ and the induced gap is small compared to $\Delta$. For $\omega\ll\Delta$, both the induced pairing term and the quasiparticle weight become independent of $\omega$. Then, the subgap spectrum of the wire can be obtained from an effective Hamiltonian, provided that the induced gap is sufficiently small. In the context of the model studied in this paper, this latter condition is guaranteed by the spin polarization of the wire. 

The renormalizations of the Hamiltonian parameters are due to the quasiparticle weight. As the coupling between wire and superconductor increases, the quasiparticle weight of the wire Green function is progressively reduced. This renormalization is directly mirrored in factors involving $4W+\xi_N|t_\uparrow|^2$ in the semiclassical approach of this paper. Such factors are involved in the semiclassical expressions \eqref{eq:vxasymp} and \eqref{eq:egap} for the Fermi velocity and the induced gap of the normal metal, respectively. Correspondingly, both quantities involve renormalizations by the quasiparticle weight in the Green function approach. At the same time, the quasiparticle weight drops out from the localization length of the Majorana bound state (or, equivalently, the coherence length of the induced superconductivity) since it is the ratio of Fermi velocity and induced gap. Again, this is consistent with our semiclassical approach which also does not involve a factor  $4W+\xi_N|t_\uparrow|^2$ in Eqs. \eqref{eq:rhohelow1} and \eqref{eq:rhohelow2}. Note that despite this absence of renormalization, the Majorana localization length depends on the bare system parameters in a nontrivial way, as it is independent of the gap of the proximity providing superconductor (see also \cite{akhmerov17}).

We finally note that our analysis excluded the presence of disorder which may or may not affect the properties of the topological phase. As discussed earlier \cite{Kupferschmidt_2011, duckheim11}, for a mean-free path $\ell$ much larger than the microscopic length scales,  the single reflection amplitude $\rho_{\rm he} \delta L$ is not affected since it is obtained by matching the wavefunctions at the short scale of the half-metal - superconductor interface. In contrast, the derivation of the reflection amplitude $r_{\rm he}^{\rm eff}$ includes multiple scattering processes at a length scale $1/|\rhoEff|$. In the absence of disorder, these add coherently to $r_{\rm he}^{\rm eff}$ because $k_x$ is conserved. Including disorder with $\ell \ll 1/|\rhoEff|$ leads to contributions from different $k_x$ for different scattering paths. Additionally, based on symmetry arguments it can be shown that $r_{\rm he}$ is anti-symmetric in $k_x$ \cite{Kupferschmidt_2011}. Hence the sum over the different paths is incoherent and there is no guarantee that $r_{\rm he}^{\rm eff}$ is unaffected by disorder. However, if $\ell \gg 1/|\rhoEff|$ the amplitudes still add coherently, and disorder is expected to not play a role. Since $1/|\rhoEff|$ is strongly decreased for a highly transparent interface, we conclude that high transparencies lead to a better protection from disorder for the Majorana bound states.

\begin{acknowledgements}
We thank Christian Kl\"ockner and Max Geier for discussions. Financial support was provided by the Institute ``Quantum Phenomenon in Novel Materials'' at the Helmholtz Zentrum Berlin, and the Deutsche Forschungsgemeinschaft (project C03 of the CRC 183).
\end{acknowledgements}

\appendix
\renewcommand\thefigure{\thesection.\arabic{figure}}   
\setcounter{figure}{0} 

\section{Cylindrical wire}

\begin{figure}
	\includegraphics[width=1\columnwidth]{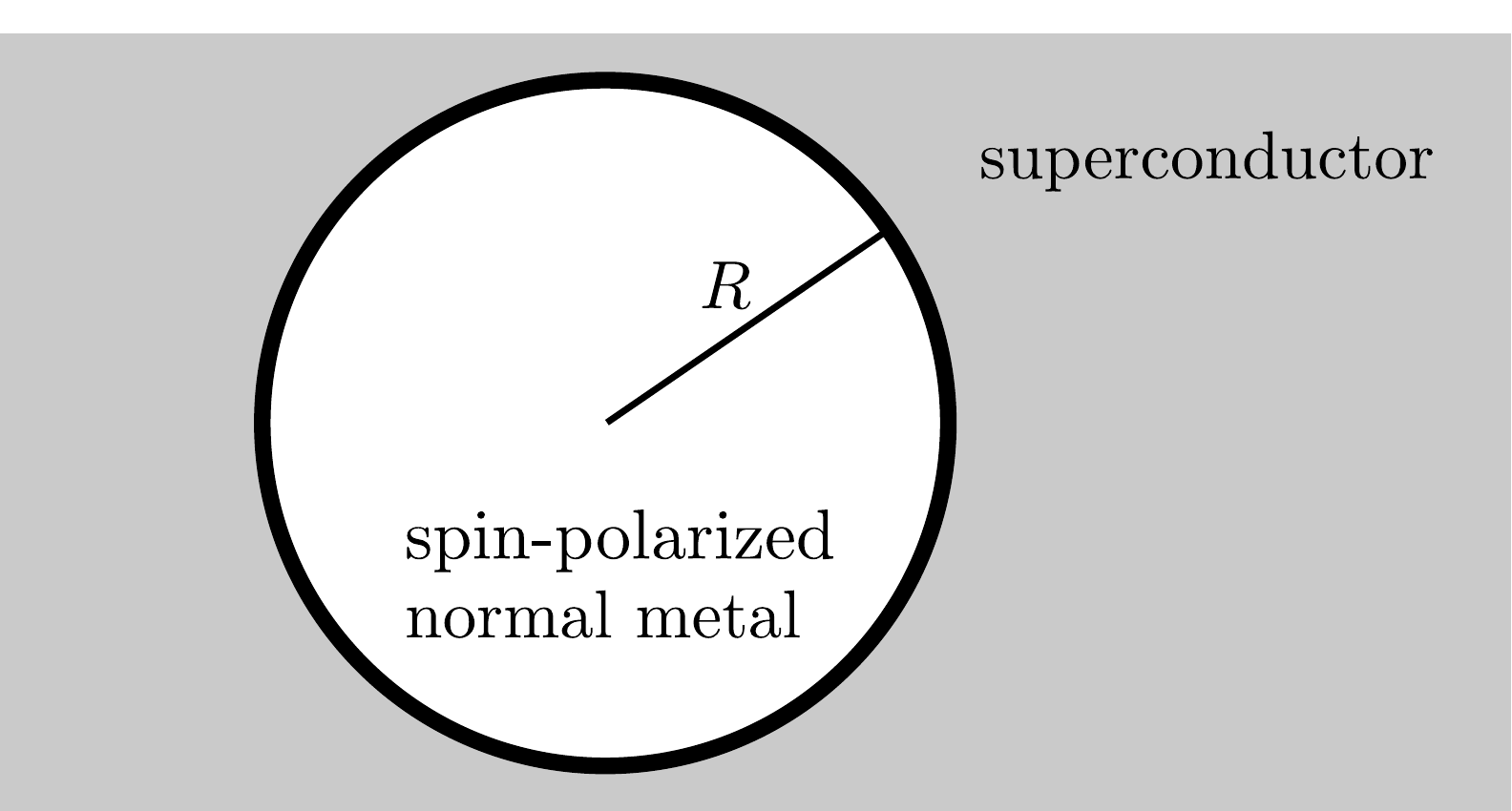}
	\caption{\label{fig:cyl_setup}
	Cross section of the cylindrical setup. A normal metal wire of radius $R$ is surrounded by an $s$-wave superconductor. Spin-orbit coupling may be present in both materials.
		}
\end{figure}

The restriction to a planar model made in the main text, in principle, accounts only for a small subset of realizable materials.  This motivates us to investigate a three dimensional analogue of the planar setup discussed in the main-text. We consider a cylindrical, half-metallic wire of radius $R$, surrounded by an $s$-wave superconductor with spin-orbit coupling in either of the two materials. A cross section of the setup is shown in Fig.\ \ref{fig:cyl_setup}. The main differences to the planar model are a change in the basis of the transverse components, from plane waves to Bessel functions, and the addition of an angular momentum quantum number. We consider only the Zeeman term induced by the magnetic field and neglect the orbital term.


Within this model and in the regime of a single transverse mode inside the wire, we show that the renormalization of the low-energy dispersion, the Majorana decay length, and the induced minigap shows essentially the same dependence on the model parameters as in the case of a planar model.


The outline of our approach is similar to the one for the planar setup. After defining the cylindrical model, first, we derive the transmission and reflection amplitudes at a normal-metal---normal-metal interface. Next, we include a finite superconducting order parameter $\Delta$, and follow the lines of the main-text in order to obtain the renormalized dispersion. Building on these results, we then derive the Majorana decay length and the minigap in the presence of the spin-orbit coupling.

\subsection{Model}

The three dimensional setup is described by the same Hamiltonian as in Eq. (\ref{eq:hamiltonian_full}), with three changes: First, we generalize to cylindrical coordinates ($r$, $\varphi$, $x$), with $x$ parallel to the wire, and radius $r \geq 0$ and angle $\varphi$ in the transverse directions. The explicit dependence on $z$ is changed according to $z\rightarrow r-R$, where $R$ is the radius of the wire. Second, we set $V_{\rm conf}(r) = 0$ for all $r$, as the cylindrical normal wire has a boundary with the superconductor only, while for the planar setup a termination at $z = -W$ was necessary. Third, we take into account the cylindrical geometry in the spin-orbit coupling tensor, such that the components along the unit vectors $\hat{\bf e}_r$, $\hat{\bf e}_\varphi$ and $\hat{\bf e}_x$ are constant. To this end, we redefine
\begin{equation}\label{eq:cyl_so_matrix_elements}
	{\bf \Omega}_{{\rm X} j} = 
	\Omega_{{\rm X} j r} \hat{\bf e}_r +
	\Omega_{{\rm X} j \varphi} \hat{\bf e}_\varphi + 
	\Omega_{{\rm X} j x} \hat{\bf e}_x
\end{equation}
where ${\rm X} = {\rm S,\,N}$.

\subsection{Cylindrical normal-normal interface}

We start our calculation by deriving the scattering amplitudes at the interface for $\Delta = 0$ and in the absence of spin-orbit coupling. In this case, the wave functions read
\begin{equation}
	\Psi_{k_x,\,m}(\vr, \varepsilon) = e^{i m \varphi + i k_x x} \psi_{k_x,\,m}(r, \varepsilon).
\end{equation}
Here we introduced the integer angular momentum quantum number $m$ and the longitudinal momentum $k_x$. In order to distinguish the quantum number $m$ from the mass of the electrons, we rename the latter to $\me$ in this appendix. The radial component, normalized to unit flux, reads
\begin{widetext}
\begin{align}\label{eq:CYL_wave_function_normal0}
	\psi_{k_x,\,m}(r, \varepsilon) =
		\sqrt{\frac{\pi \me}{2 \hbar}}
		\times
		\begin{pmatrix}
			\ceU \Hone_m (\kUr(\varepsilon) r) + \ceU' \Htwo_m (\kUr(\varepsilon) r) \\
			0\\
			\chU \Htwo_m (\kUr(-\varepsilon) r)+ \chU'  \Hone_m (\kUr(-\varepsilon) r) \\
			0
		\end{pmatrix}
		&+
		\sqrt{\frac{2 \pi \me}{\hbar}}
		\begin{pmatrix}
			0\\
			\ceD I_m (\kaDr(\varepsilon) r) \\
			0 \\
			\chD I_m (\kaDr(-\varepsilon) r)
		\end{pmatrix}
\end{align}
\end{widetext}
for $r<R$ and
\begin{align}\label{eq:CYL_wave_function_SC1}
	\psi_{k_x,\,m}(r) &=
		\sqrt{\frac{\pi \me}{2 \hbar}}
		\\ \nonumber
		&\times
		\begin{pmatrix}
			\deU \Htwo_m (\kSr r) + \deU' \Hone_m (\kSr r) \\
			\deD \Htwo_m (\kSr r) + \deD' \Hone_m (\kSr r) \\
			\dhU \Hone_m (\kSr r) + \dhU'  \Htwo_m (\kSr r) \\
			\dhD \Hone_m (\kSr r) + \dhD'  \Htwo_m (\kSr r) 
		\end{pmatrix}
\end{align}
for $r > R$.
Here, $H_{m}^{(1,\,2)}$ are the Hankel functions of first and second kind and $I_m$ is the modified Bessel function of the first kind. We drop the $m$-indices of the $c$ and $d$ coefficients for the sake of compactness.
The wave and decay numbers are
\begin{align}
	\kUr(\varepsilon) = & \sqrt{\kFU^2 - k_x^2 + 2 \me \varepsilon/\hbar^2}
	, \\
	\kaDr(\varepsilon) = & \sqrt{\kaFD^2 + k_x^2 - 2 \me \varepsilon/\hbar^2}
	, \\
	\kSr = & \sqrt{\kFS^2 - k_x^2}.
\end{align}
Here we neglected the $\varepsilon$-dependence in $\kSr$ because we will apply the Andreev approximation for $r > R$ in the next section.

The $c$-coefficients are constrained by the requirement that the wavefunction has to be well-behaved at the origin, which is satisfied if the Hankel-functions add up to the Bessel functions of the first kind $J_m(z) = [H^{(1)}_m(z) + H^{(2)}_m(z)]/2$. This corresponds to fixing $\ceU' = \ceU$ and $\chU' = \chU$. No conditions on $\ceD$ and $\chD$ are required, as $I_m$ is well-behaved at the origin.

The relations between the $c$- and $d$-coefficients are determined by continuity of the wavefunction at the interface, and by
\begin{equation}
	\psi_{k_x, m} '(R + \delta, \varepsilon) = \psi_{k_x, m}'(R - \delta, \varepsilon) + 2 \omega \psi_{k_x, m}(R, \varepsilon)
\end{equation}
with $\delta \rightarrow 0$. Solving the matching conditions relates the in- and out-going modes by
\begin{align} \label{eq:coeff_realtions_Delta0_maj}
	\nonumber
	\begin{pmatrix}
		\deU' \\ \ceU' \\ \dhU' \\ \chU'
	\end{pmatrix} = &
		\begin{pmatrix}
			t_{\uparrow m}(\varepsilon) & r'_{\uparrow m}(\varepsilon) & 0 & 0 
			\\
			r_{\uparrow m}(\varepsilon) & t_{\uparrow m}(\varepsilon) & 0 & 0
			\\
			0 & 0 & t^*_{\uparrow m}(-\varepsilon) & r'^*_{\uparrow m}(-\varepsilon)
			\\
			0 & 0 & r^*_{\uparrow m}(-\varepsilon) & t^*_{\uparrow m}(-\varepsilon)
		\end{pmatrix}
		\begin{pmatrix}
			\ceU \\ \deU \\ \chU \\ \dhU
		\end{pmatrix},
	\\ \\
	\begin{pmatrix}
		\deD' \\ \ceD \\ \dhD' \\ \chD
	\end{pmatrix} = &
		\begin{pmatrix}
			r'_{\downarrow m}(\varepsilon) & 0 \\
			t_{\downarrow m}(\varepsilon) & 0 \\
			0 & r_{\downarrow m}^{\prime *}(-\varepsilon) \\
			0 & t_{\downarrow m}^*(-\varepsilon)			
		\end{pmatrix}
		\begin{pmatrix}
			\deD \\ \dhD
		\end{pmatrix}.\label{eq:coeff_realtions_Delta0_min}
\end{align}
Here, we dropped the dependence on $k_x$ in order to keep the notation compact. Note that Eqs \eqref{eq:coeff_realtions_Delta0_maj} and \eqref{eq:coeff_realtions_Delta0_min} are identical to the ones for a planar setup \cite{Kupferschmidt_2011}, while the parametrization of the transmission ($t$) and reflection ($r$) amplitudes differs.

By applying the matching conditions at the interface, $r=R$,  we obtain
\begin{widetext}
\begin{align}
	t_{\uparrow m}(k_x, \varepsilon) &= \frac{4i/\pi R}
   		{
   			\kSr \Hone_{m-1} (\kSr R) \Htwo_{m}(\kUr R)
   			- \Hone_m (\kSr R)
   			\left[
   				 k_{\uparrow r} \Htwo_{m-1}(\kUr R)
   				+ 2 \omega  \Htwo_{m}(\kUr R)
   			\right]
   		}
   	,\label{eq:cyl_amp_t}
   	\\
	r_{\uparrow m}(k_x, \varepsilon) &= 
		\frac{ - \HpUR{m} + t_{\uparrow m}  \HpSR{m}}{ \HmUR{m} } 
	,
	\\
	r'_{\uparrow m}(k_x, \varepsilon) &= 
		\frac{ - \HmSR{m} + t_{\uparrow m} \HmUR{m} }{ \HpSR{m} } 
	,
	\\
	t_{\downarrow m}(k_x, \varepsilon) &= \frac{2 i/ \pi R}
		{
			\kSr  \Hone_{m-1} (\kSr R) \IDR{m}
			- \Hone_m (\kSr R) \left[
				\kaDr \IDR{m-1}
				+ 2 \omega  \IDR{m} 
   			\right]
   	}
   	,
   	\\
   	\label{eq:cyl_amp_rpD}
	r'_{\downarrow m}(k_x, \varepsilon) &= 
		e^{i \varphi_{\downarrow m}(k_x, \varepsilon)} 
		= \frac{-\HmSR{m} + 2 t_{\downarrow m} \IDR{m}}{\HpSR{m}}
	.
\end{align} 
\end{widetext}
Here we dropped the dependencies on $(k_x, +\varepsilon)$ on the right-hand side.

It is useful to consider limiting cases of the transmission and reflection amplitudes. We assume $\kSr \gg \kUr,\ \kaDr$. First, in the limit where $m$ is small compared to the arguments of the Bessel functions, the amplitudes are related to their counterparts in the planar model [see Eqs. \eqref{eq:2d_tUp} to \eqref{eq:2d_rpUp}, \eqref{eq:2d_rpDown} and \eqref{eq:2d_tauDown}] by
\begin{align}
	t_{\uparrow m}(k_x, 0) &= e^{i (\kUr - \kSr) R} t_{\uparrow}(k_x)|_{z\rightarrow r}
   	, 
   	\\
	r_{\uparrow m}(k_x, 0) &= - i (-1)^m e^{2i \kUr R} r_{\uparrow}(k_x)|_{z\rightarrow r}
	,
	\\
	r'_{\uparrow m}(k_x, 0) &= i (-1)^m e^{-2 i \kSr R}  r_{\uparrow}'(k_x)|_{z\rightarrow r}
	,
	\\ \label{eq:cyl_largeR_phiD}
	r'_{\downarrow m}(k_x, 0) &= i (-1)^m  e^{-2 i \kSr R} r_{\downarrow}'(k_x)|_{z\rightarrow r}
	,
	\\
	t_{\downarrow m}(k_x, \varepsilon) &= 
			e^{(2 m + 1) \pi/4 - \kaDr R - i \kSr R}
			\left. 			
			\sqrt{\frac{\kaDr}{\kUr}} \tau_{\downarrow}(k_x)
			\right|_{z\rightarrow r}
	.
\end{align}

Next, we take the limit $\kSr R \ll |m|$. Consider the case $R=0$ (no wire). Then the radial components need to be Bessel functions of the first kind, which vanishes for $\kSr r \ll m$, and the overlap with the wire will stay negligible for finite $R$. Thus even for finite $R$, large-$m$ modes have a vanishing overlap with the wire and $r'_{\uparrow m} = r'_{\downarrow m} = r_{\uparrow m}=1$ and $t_{\uparrow m} = 0$. 

Finally, for intermediate $m$ where $\kUr R$, $\kaDr R \ll |m| \ll \kSr R$, we find $r_{\uparrow m} = 1$ and $t_{\uparrow m} = 0$ to lowest order. Consequently, the radial components become small for $r < R$. The remaining two amplitudes $r_{\uparrow m}'$ and $r_{\downarrow m}'$ are both of magnitude one, with their phases depending on $m$, $\kSr$, $\kaDr$ and $w$.

In the case of an ideal interface, $w = 0$ and $\kFU = \kFS$, the amplitudes for the majority carriers reduce to $t_{\uparrow m} = 1$ and $r_{\uparrow m} = r'_{\uparrow m} = 0$. This can be verified by using the Wronskian of the Hankel functions and the unitarity of the scattering matrix \cite{nist10}.

\subsection{Renormalization of the Fermi velocity} \label{sec:appendix_cyl_dispersion}

Next we include a finite superconducting order parameter $\Delta$. As described in section \ref{sec:renormalization} of the main text, this is expected to confine excitations with energies $\varepsilon < \Delta$ to the normal region, $r < R$, with evanescent components in the superconducting region that decay at a length scale of order of the coherence length $\xi_\varepsilon$. The additional weight in the superconductor, as well as the change of the matching conditions at the boundary lead to a renormalization of the wire dispersion. 

Following the lines of the main text, we can derive this renormalization by first considering the majority wavefunction for $r < R$. It reads
\begin{align}\label{eq:CYL_wave_function_normal1}
	\psi_{k_x, \,m}(r, \varepsilon) =&\,
			\begin{pmatrix}
			u_{\uparrow,k_x,m}(r, \varepsilon) \\
			0\\
			0 \\
			v_{\downarrow,k_x, m}(r, \varepsilon)
		\end{pmatrix},
	\\ \nonumber
	u_{\uparrow,k_x,m}(r, \varepsilon) =&\,
		\sqrt{\frac{\pi \me}{2 \hbar}}
		\left[
			\Hone_m (\kUr(\varepsilon) r) 
		\right.
		\\ & \left.
			+ \reeC(k_x, \varepsilon) \Htwo_m (\kUr(\varepsilon) r)
		\right],
	\\
	v_{\downarrow,k_x,m}(r, \varepsilon) =&\,
		\sqrt{\frac{2 \pi \me}{\hbar}}\chD I_m (\kaDr(-\varepsilon) r).
\end{align}

The amplitude $\reeC$ is derived by applying wavefunction matching at $r = R$ and by requiring decaying modes for $r \rightarrow \infty$. By applying the latter condition we obtain the wavefunction inside the superconductor, which reads
\begin{align}\label{eq:CYL_wave_function_SC2}
	\psi_{k_x,\,m}(r, \varepsilon) &=
		\sqrt{\frac{\pi \me}{2 \hbar}}
		\times  
		\\ \nonumber
		& \left[ 
			\frac{			
					\Hone_m \left(\kSr r + i \frac{r}{\xi_\varepsilon} \right)
			}{
					A_m^{(1)}
			}
		 	\begin{pmatrix}
				d'_{\uparrow}\\
				0 \\
				0 \\
				d'_{\uparrow}e^{-i \eta - i \phi}
			\end{pmatrix}
		\right.		
		\\ \nonumber	
		&+
		\left.
			\frac{
					\Htwo_m \left(\kSr r - i \frac{r}{\xi_\varepsilon} \right)
			}{
					A_m^{(2)}
			}
			\begin{pmatrix}
				d_{\uparrow} \\
				0 \\
				0 \\
				d_{\uparrow}e^{i \eta - i \phi}
			\end{pmatrix}
		\right],
\end{align}
where the factors
\begin{equation}
	A_m^{(1/2)} = \frac{
		H_m^{(1/2)} \left(\kSr R \pm i \frac{R}{\xi_\varepsilon} \right)
	}{
		H_m^{(1/2)} \left(\kSr R \right)
	}
\end{equation}
ensure that for $m \lesssim \kSr R$, the exponential decay of the Hankel function, proportional to $e^{-r/\xi_\varepsilon}$, is canceled at $r=R$. We dropped the $\varepsilon$ and $k_x$ dependencies for the sake of compactness.
Within Andreev approximation, $\hbar^2 \kSr^2/2 \me \ll \Delta$, the interface can be treated as an interface between two normal metals. Thus, the $c$ and $d$ coefficients for the superconductor and wire components are related by Eqs. \eqref{eq:coeff_realtions_Delta0_maj} and \eqref{eq:coeff_realtions_Delta0_min}. Combining the interface matching relations with Eq. \eqref{eq:CYL_wave_function_SC2} and setting $\ceU = 1$ yields 
\begin{align}\label{eq:ree_cyl}
\nonumber
\reeC(k_x, \varepsilon) &=  
	r_{\uparrow m}(k_x, \varepsilon) 
	\\ & \ \ \
	+ \frac{
		t_{\uparrow m}(k_x, \varepsilon)^2 
	}{
		r'_{\downarrow m}(k_x, -\varepsilon)e^{2 i \etaE} 
		- r'_{\uparrow m}(k_x, \varepsilon)
	},
	\\
	d_{\uparrow}(k_x, \varepsilon) &= 
		\frac{
			t_{\uparrow m}(k_x, \varepsilon)
		}{
			r'_{\downarrow m}(k_x, -\varepsilon)e^{2 i \etaE} 
		- r'_{\uparrow m}(k_x, \varepsilon)
		},		
	\\
	d'_{\uparrow}(k_x, \varepsilon) &= r'_{\downarrow m}(k_x, -\varepsilon) e^{2 i \etaE} d_{\uparrow}(k_x, \varepsilon),
	\\ \label{eq:cyl_coeff_h_downarrow}
	\chD(k_x, \varepsilon) &= t^*_{\downarrow m}(k_x, -\varepsilon)  e^{-i \etaE -i \phi} d'_{\uparrow}(k_x, \varepsilon).
\end{align}
Equations \eqref{eq:ree_cyl} to \eqref{eq:cyl_coeff_h_downarrow} are identical to the ones in a planar setup at $\varepsilon=0$ in terms of the interface amplitudes \cite{Kupferschmidt_2011}, while the parametrization of the amplitudes is different.


The requirement of the wavefunction being well behaved at $r=0$ restricts the normal reflection amplitude by
\begin{equation}\label{eq:cyl_non_linear_ev_equation}
	1 = \reeC(k_x, \varepsilon_m).
\end{equation}
Solving this equation yields the dispersion $\varepsilon_m(k_x)$ and the renormalized velocity $v_{x,m}(\varepsilon) = |d\varepsilon_m/dk_x|/\hbar$. In the following, we provide limiting solutions to Eq. \eqref{eq:cyl_non_linear_ev_equation} for unit transparency, as well as low transparency.

\begin{figure}
	\includegraphics[width=1\columnwidth]{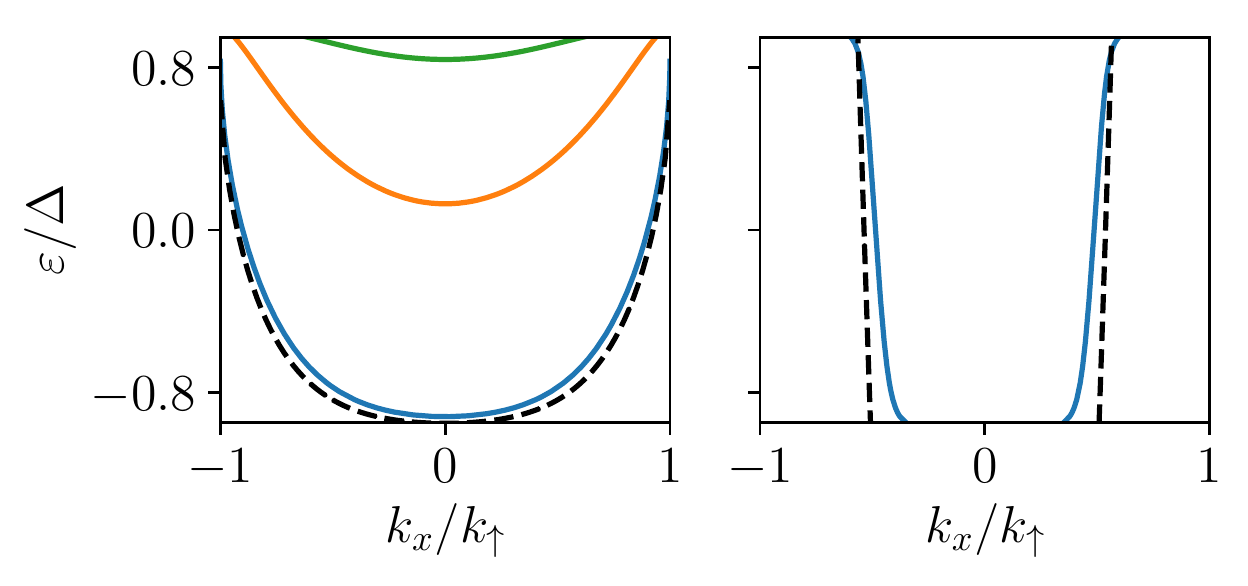}
	\caption{\label{fig:dispersion_cyl}
	Dispersion for the cylindrical setup with different interface transparencies. We choose $\kFS = \kFU$ and $w =0$ on the left, which yields $t_{\uparrow m} =1$, and $w/\kFU = 2$ with $\kFS=\kFU$ on the right, corresponding to $|t_{\uparrow 0}|^2 = 0.2$ for perpendicular incidence. The solid lines show the numerical solution of Eq. \eqref{eq:cyl_non_linear_ev_equation}, with angular momentum numbers $m=0$ (blue), $m=\pm1$ (orange) and $m = \pm 2$ (green). In the right plot, we find in-gap solutions for $m=0$ only. The dashed lines shows the predictions from Eq. \eqref{eq:cyl_dispersion_T1}(left) and \eqref{eq:cyl_eps_low_transparency}(right). The remaining parameters are $\kFU R = 0.8 \pi$, $\kaFD/\kFU = 2$ and $(\hbar \pi/R)^2/2 \me \Delta = 50$.
		}
\end{figure}

\begin{figure}
	\includegraphics[width=1\columnwidth]{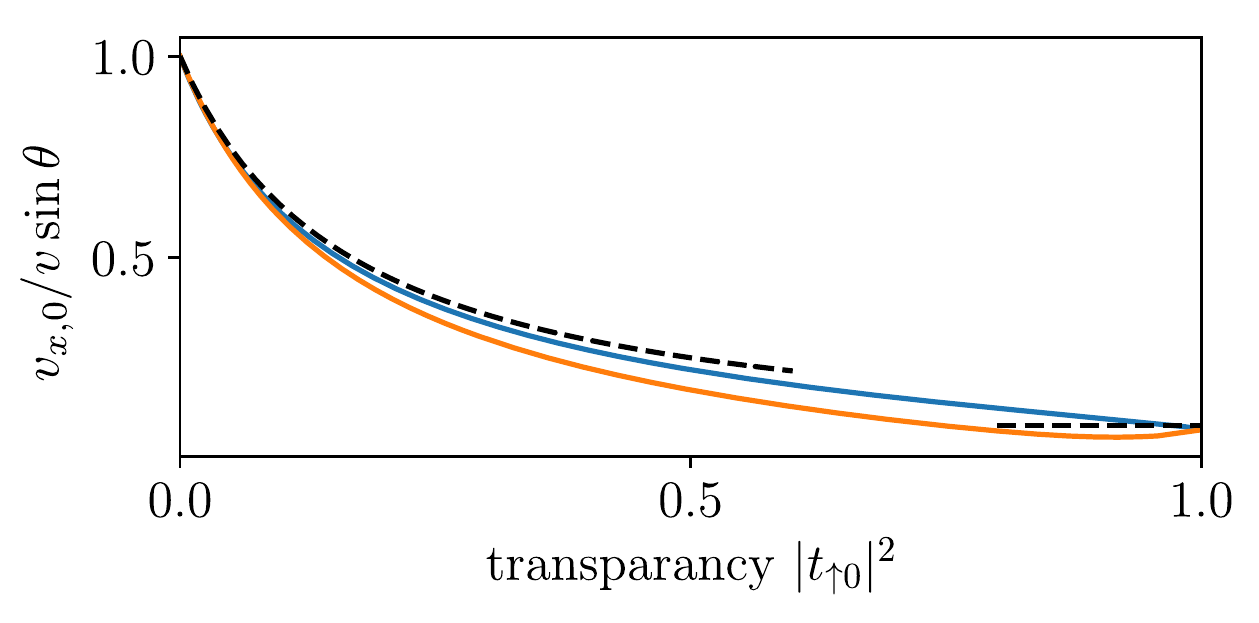}
	\caption{\label{fig:velocity_cyl}
		Velocity renormalization as a function of transparency on a semi-logarithmic scale.  For the blue (upper) line, we tune the transparency by varying $w$ while keeping $\kFU = \kFS$ fixed. For the orange (lower) line, we vary $\kFS/\kFU \geq 1$ with $w =0$ fixed.  The dashed lines show the predictions for $|t_{\uparrow 0}|^2 \ll 1$ and for $|t_{\uparrow 0}|^2 = 1$.  The remaining parameters are the same as in Fig. \ref{fig:dispersion_cyl}.
	}
\end{figure}

For a transparent interface, $w = 0$ and $\kFS = \kFU$, Eq. \eqref{eq:cyl_non_linear_ev_equation} reduces to 
\begin{equation}
	2 \eta(\varepsilon_m) + \varphi_{\downarrow m}(k_x, \varepsilon_m) = 2 \pi n,
\end{equation}
with an integer number $n$. Within Andreev approximation inside the wire, $\hbar^2 \kUr^2/2 \me \Delta \ll 1$ and for $\hbar^2 \kaDr^2/2 \me \Delta \ll 1$, the energy dependence in $\varphi_{\downarrow m}$ can be neglected, and we obtain
\begin{equation}
	\varepsilon_m(k_x) = \pm \Delta \cos \varphi_{\downarrow m}(k_x).
\end{equation}
For $m \gg \kUr R$, the phase $\varphi_{\downarrow m}$ vanishes and $\varepsilon_m = \pm \Delta$. Hence, the large $m$ modes are gapped out.
For $\kUr R \gtrsim m$, we can use the approximation \eqref{eq:cyl_largeR_phiD} to obtain 
\begin{equation}
	2 \kUr R = 2 \eta(\varepsilon_m) + \varphi_{\downarrow}(k_x) + (4 n + 2 m + 1)\pi/2,
\end{equation}
where $\varphi_\downarrow(k_x)$ is defined in Eq. \eqref{eq:2d_rpDown}. Solving for $\varepsilon_m$, we get
\begin{equation}\label{eq:cyl_dispersion_T1}
	\varepsilon_m(k_x) = \pm \Delta \cos \left[ 
	\kUr R  - \frac{\varphi_{\downarrow}(k_x)}{2}
	- (2 m + 1 ) \frac{\pi}{4}	
	\right],
\end{equation}The velocity is obtained by taking the derivative,
\begin{equation}\label{eq:cyl_velocity_T1}
	v_{x, m}(k_x) = \sqrt{\Delta^2 - \varepsilon_m^2} \frac{k_x R}{\hbar \kUr} 
	\left|1 - \frac{1}{\kaDr R} \right|,
\end{equation}
which is identical to the one for the planar model upon replacing $r$ by $z$ and $R$ by $W$. The analytical predictions in Eqs. \eqref{eq:cyl_dispersion_T1} and \eqref{eq:cyl_velocity_T1} are compared to a direct numerical solution of Eq. \eqref{eq:cyl_non_linear_ev_equation} in Figs. \ref{fig:dispersion_cyl} and \ref{fig:velocity_cyl}. Both limiting cases show good agreement.

Next, we consider the limit of $w$, $\kSr \gg \kUr, \kaDr$. For $m \gg \kSr R$ the overlap with the wire vanishes and all modes are gapped out, $\varepsilon = \pm \Delta$. For $m \lesssim \kSr R$, we obtain
\begin{widetext}
\begin{equation}
\reeC(k_x, \varepsilon) = -\frac{ H_{m}^{(1)}(\kUr R) }{ H_{m}^{(2)}(\kUr R)}
	\frac{
		|H_{m}^{(1)}(\kUr R)|^2 \pi R \Delta (\kSr^2 + 4 w^2) + 2 i \kUr (\kSr \varepsilon/\Delta + 2 w)	
	}{
		|H_{m}^{(1)}(\kUr R)|^2 \pi R \Delta (\kSr^2 + 4 w^2) - 2 i \kUr (\kSr \varepsilon/\Delta + 2 w )
	}.
\end{equation}
\end{widetext}
Equation \eqref{eq:cyl_non_linear_ev_equation} enforces $\reeC = 1$ and to zeroth order in $\kUr/w$ and $\kUr/\kSr$ we get $H_{m}^{(1)}(\kUr R) + H_{m}^{(2)}(\kUr R) =0$. The solutions of this equation correspond to the zeros of the $m$-th order Bessel function of the first kind. For positive $\kUr$, this prohibits solutions with $m \gg \kUr R$, which allows us to use the small $m$ approximation. To leading order in $\kUr/w$ and $\kUr/\kSr$ we get
\begin{align} \label{eq:cyl_reeC_low_transparancy}
\reeC(k_x, \varepsilon) \approx&\, -e^{2 i \kUr R - i ( 2 m + 1) \frac \pi 2}
	\\ &\times	\nonumber
	\frac{
		\kSr^2 + 4 w^2 + i \kUr (\kSr \varepsilon/\Delta + 2 w)	
	}{
		\kSr^2 + 4 w^2 - i \kUr (\kSr \varepsilon/\Delta + 2 w )
	}.
\end{align}
In the following, we will focus on the regime where $\kUr R$ is of order one and small enough such that only a single solution exists for low energies. This solution will have $m = 0$, which allows us to transform Eq. \eqref{eq:cyl_reeC_low_transparancy}  to
\begin{equation}\label{eq:cyl_kz_lowT_quantization}
	\kUr = 
		\frac{3 \pi}{4 R}
		\left[
			1 -
			\frac{ \kSr\varepsilon_0/\Delta + 2 w}
			{R (\kSr^2 + 4 w^2)}
		\right],
\end{equation}
which yields
\begin{align}\label{eq:cyl_eps_low_transparency}
	\varepsilon_0(k_x) =&\, \frac{\hbar^2}{2 \me} 
	\left[ 
		k_x^2 - \kFU^2  + \frac{9\pi^2}{16 R^2}
		\left(
			1 -
			2\frac{\kSr \varepsilon_0 /\Delta + 2 w}
			{R (\kSr^2 + 4 w^2)}
		\right)
	\right].
\end{align}
The $\varepsilon$-dependence on the right hand side can be neglected, yielding an explicit equation for $\varepsilon$.
Taking the derivative with respect to $k_x$ results in
\begin{equation}\label{eq:cyl_vx_low_transparency}
	v_{x, 0} = \frac{v_r \sin \theta_r }{1 + \xi_{{\rm N}0} |t_{\uparrow 0}|^2 / 4 R},
\end{equation}
where $v_r = \hbar \kFU/\me$, $\sin \theta_r = k_x/\kFU$ and $\xi_{{\rm N}0} = \hbar^2 \kUr / \me \Delta$.
The renormalization of the dispersion that is present in Eqs. \eqref{eq:cyl_eps_low_transparency} and \eqref{eq:cyl_vx_low_transparency} is the same as the one for the planar setup, see Eqs. \eqref{eq:2d_eps_low_transparency} and \eqref{eq:vxasymp}, up to the change $W \rightarrow R$, $z \rightarrow r$ and the factor $9/16$ in Eq. \eqref{eq:cyl_eps_low_transparency}, which originates in the basis change from plane waves for two dimensions to Bessel functions in three dimensions. The approximations in Eqs. \eqref{eq:cyl_eps_low_transparency} and \eqref{eq:cyl_vx_low_transparency} are shown as the dashed line in Figs. \ref{fig:dispersion_cyl} and \ref{fig:velocity_cyl}. They show good agreement for small transparencies. 

Fig. \ref{fig:dispersion_cyl} also shows that higher angular momentum modes are gapped out, and that they penetrate deeper into the gap in the high transparency case. The renormalization of the velocity for the $m=0$ mode is qualitatively the same as the renormalization in the planar setup, for all transparencies.

\subsection{Effects of spin-orbit coupling}

Spin-orbit coupling is expected to have the same effects as in the planar-model, making Andreev reflection between majority spin electrons and majority spin holes possible, opening a minigap $\varepsilon_{\rm gap}$ and allowing Majorana bound states to form at the end of the cylindrical wire. 

We assume spin-orbit coupling to be weak, such that we can treat its effects within first order perturbation theory and neglect finite-energy corrections of order $\varepsilon/\Delta$. Furthermore, we restrict ourselves to the single mode regime, where $m=0$.

The electron-like wave functions $\ket{\psi_{{\rm e} \pm}}$, travelling into the positive ($+$) or negative ($-$) $x$ direction, are given by 
\begin{equation}
	\psi_{{\rm e} \pm}(\vr, \varepsilon) = 
		\frac{
			\sqrt{\vUr} e^{i k_x(\varepsilon) x}
		}{
			\sqrt{2 \pi v_{x, 0} \mathcal{N}_0}
		}
		\psi_{ \pm k_x(0), 0}(r, \varepsilon)
\end{equation}
with $\psi_{ \pm k_x(\varepsilon), 0}(r, \varepsilon)$ defined in Eqs. \eqref{eq:CYL_wave_function_normal1} and \eqref{eq:CYL_wave_function_SC2} and
\begin{equation}
	k_x(\varepsilon) = \sqrt{\kFU^2 - \kUr^2} + \frac{\varepsilon}{\hbar v_{x,0}}.
\end{equation}
The velocity $v_{x,0}$ and $\kUr$ are taken from the calculation of the dispersion in Sec. \ref{sec:appendix_cyl_dispersion}.
The normalization constant is obtained by normalizing to unit flux along the wire. It reads
\begin{equation}\label{eq:cyl_normalization_constant}
	\mathcal{N}_0 = 2 R 
	+2 \xi_{\rm N} \frac{ |t_{\uparrow 0}|^2}{ |r'_{\downarrow 0} + r'_{\uparrow 0}|^2},
\end{equation}
where we defined $\xi_{\rm N} = \hbar^2 \kUr/\me \Delta$, neglected the minority spin contribution in the wire and expanded the Bessel functions in terms of plane waves. 
The renormalization present in Eq. \eqref{eq:cyl_normalization_constant} is similar to the one in the planar setup, see Eq. \eqref{eq:2d_normalization_constant}.

The hole-like wave functions $\ket{\psi_{{\rm h} \pm}}$, travelling into positive ($+$) or negative($-$) $x$-direction, are obtained by applying particle-hole symmetry
\begin{equation}
	\psi_{{\rm h} \pm}(\vr, \varepsilon) = \tau_x \left[\psi_{{\rm e} \pm}(\vr, -\varepsilon) \right]^*.
\end{equation}

In order to study how spin-orbit coupling changes these states, we consider a segment $0 < x < \delta L$, in which spin-orbit coupling is turned on while it is zero elsewhere. For sufficiently small $\delta L$, the reflection amplitude becomes linear in $\delta L$ and is given by the matrix element \eqref{eq:matrix_element}, with the spin-orbit coupling tensor defined according to Eq. \eqref{eq:cyl_so_matrix_elements}. Evaluating the matrix element in the single mode limit and for $\kaFD R \gtrsim 1$ yields
\begin{widetext}
\begin{align}\label{eq:cyl_rho_he}
	\rho_{\rm he, c} =&\, 
		\frac{
			- \kUr \hbar k_x (\OmSxx + i \OmSyx)
			e^{-i \phi + 2 i \kUr R}
			t_{\uparrow}^2 (1 + r'^2_{\downarrow})
		}{
			v_{x,0} \mathcal{N}_0 \kSr^2 (r'_\downarrow + r'_\uparrow)^2
		}
		\\ \nonumber
		& -
		\frac{
			2 \hbar k_x (\OmNxx + i \OmNyx) e^{-i \phi} \tau_\downarrow t_\uparrow
		}{
			v_{x,0} \mathcal{N}_0 (\kUr^2 + \kaDr^2) ( r'_\downarrow + r'_\uparrow )
		}
		\left[
			\kaDr \left( 1 - i e^{2 i \kUr R} \right)
			+ i \kUr \left(1 + i e^{2 i \kUr R} \right)	
		\right],
\end{align}
\end{widetext}
where $t_\uparrow$, $\tau_\downarrow$, $r'_\uparrow$ and $r'_\downarrow$ are the interface amplitudes defined in the planar setup.

 The remaining amplitudes for reflection from the right, as well as from holes to electrons are obtained by the same symmetry arguments as the ones discussed below Eq. \eqref{eq:rho_he}. Similarly, the reflection amplitude for a segment of length $L$, as well as the gap is obtained by the same arguments as in the main text. This allows us to define the minigap
\begin{equation}
	\varepsilon_{0, \rm c} = \hbar v_{x,0} |\rho_{\rm he, c}|,
\end{equation}
and the localization length
\begin{equation}
	l_{\rm maj, c} = |\rho_{\rm he, c}|^{-1}.
\end{equation}

Equation \eqref{eq:cyl_rho_he} is almost identical to $\rho_{\rm he}$ in Eq. \eqref{eq:rho_he}. Indeed, in the single mode limit and for $\kUr R \gtrsim 1$ we have
\begin{equation}
	1 = r_{{\rm ee},0} = -i e^{2 i \kUr R} r_{\rm ee}, 
\end{equation}
with $r_{\rm ee}$ defined in Eq. \eqref{eq:ree_2d}, allowing us to identify $\rho_{\rm he} = \rho_{\rm he, c} i e^{-2 i \kUr R}$, upon replacing the labels $r$ by $z$ and $R$ by $W$. The asymptotic expansions for $\rho_{\rm he}$ in the limit $|t_{\uparrow}|^2 \ll 1$ are then obtained by replacing the factors $\pi$ in Eqs. \eqref{eq:rhohelow1} and \eqref{eq:rhohelow2} by factors of $3 \pi/4$, which originates in the difference of Eqs. \eqref{eq:2d_kz_lowT_quantization} and \eqref{eq:cyl_kz_lowT_quantization}.

\begin{figure}
	\includegraphics[width=1\columnwidth]{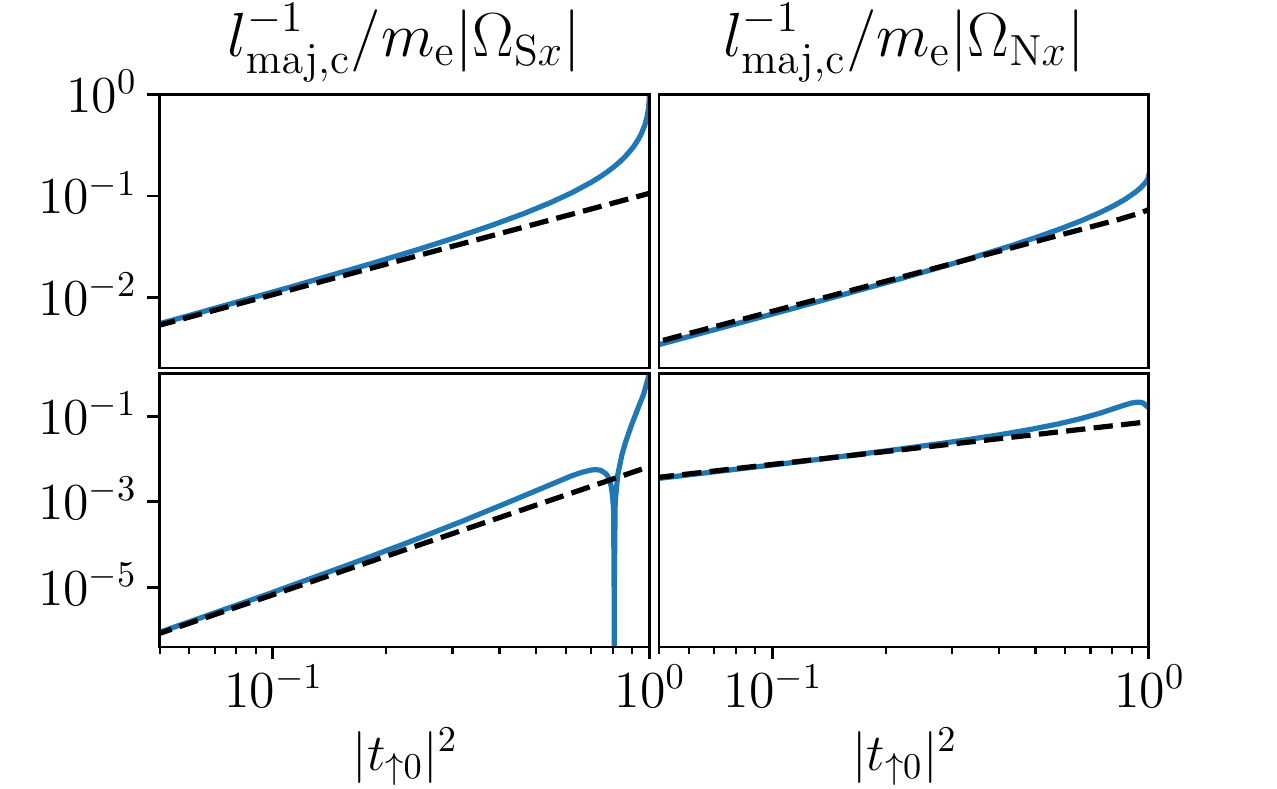}
	\caption{\label{fig:decay_cyl}
		Inverse localization length as a function of interface transparency for the cylindrical setup and the zero angular-momentum mode. We choose matched Fermi velocities (top row) and zero potential barrier $w = 0$ (bottom row), with spin-orbit coupling in the superconductor (left column) and in the normal metal (right column). The dashed curves show the weak transparency results, the solid lines are obtained by numerically solving Eq. \eqref{eq:cyl_non_linear_ev_equation} and using Eq. \eqref{eq:cyl_rho_he}. The remaining parameters are $\kUr R = 0.8 \pi$, $\kaFD/\kFU = 2$ and $(\hbar \pi/R)^2/2 \me \Delta = 50$. 
	}
\end{figure}

The low transparency approximation for the Majorana decay length is compared to a numerical solution of Eq.~\eqref{eq:cyl_non_linear_ev_equation} in Fig.~\ref{fig:decay_cyl}. Good agreement with the low transparency approximation is found for $|t_{\uparrow 0}|^2 \lesssim 0.5$. For larger transparencies deviations occur. In case of spin-orbit coupling being present in the superconductor, and no potential barrier at the interface, we find that the gap closes and reopens at transparencies close to unity (bottom left plot).

In conclusion, the velocity, decay length, and hence also the induced minigap, show essentially the same dependence on the model parameters as for the planar setup discussed in the main text.

\FloatBarrier\bibliographystyle{apsrev4-1}
\bibliography{references11}

\end{document}